\documentclass[sigconf,plain]{acmart}

\usepackage{tikz}
\usepackage{amsmath}
\usepackage{subcaption}
\usepackage{tablefootnote}
\usepackage{filecontents}
\usepackage{colortbl}
\usepackage{hhline}
\usepackage{booktabs}
\usepackage{graphicx}
\usepackage{xspace}
\usepackage{xcolor}
\usepackage{microtype}
\usepackage{listings}
\usepackage{cleveref}
\usepackage{url}
\usepackage{hyperref} 
\usepackage{breakurl}
\usepackage{multirow}
\usepackage{balance}
\usepackage{enumitem}
\usepackage[font=small,format=plain,labelfont=bf,textfont=it]{caption}

\definecolor{codegreen}{rgb}{0,0.6,0}
\definecolor{codegray}{rgb}{0.5,0.5,0.5}
\definecolor{codepurple}{rgb}{0.58,0,0.82}
\definecolor{backcolour}{rgb}{0.95,0.95,0.92}

\lstdefinestyle{mystyle}{
    backgroundcolor=\color{backcolour},   
    commentstyle=\color{codegreen},
    keywordstyle=\color{magenta},
    numberstyle=\tiny\color{codegray},
    stringstyle=\color{codepurple},
    basicstyle=\ttfamily\footnotesize,
    breakatwhitespace=false,         
    breaklines=true,                 
    captionpos=b,                    
    keepspaces=true,                 
    numbers=left,                    
    numbersep=5pt,                  
    showspaces=false,                
    showstringspaces=false,
    showtabs=false,                  
    tabsize=2
}

\definecolor{Gray}{gray}{0.9}

\hypersetup{
     colorlinks=true,
     linkcolor=black,
     filecolor=blue,
     citecolor = black,      
     urlcolor=blue
}
\renewcommand\footnotetextcopyrightpermission[1]{} 
\graphicspath{ {./figures/} }
\makeatletter
\def\input@path{{./sections/}}
\makeatother
\newcommand{\etal}{{\it et al.}}
\newcommand{\name}{nPrint\xspace}
\usepackage[symbol]{footmisc}

\title{New Directions in Automated Traffic Analysis}

\author{Jordan Holland}
\affiliation{%
    \institution{Princeton University}
    \city{Princeton}
    \state{New Jersey}
    \country{USA}
    }
    \email{jordanah@princeton.edu}

\author{Paul Schmitt}
\affiliation{%
    \institution{Princeton University}
    \city{Princeton}
    \state{New Jersey}
    \country{USA}
    }
    \email{pschmitt@cs.princeton.edu}

\author{Nick Feamster}
\affiliation{%
    \institution{University of Chicago}
    \city{Chicago}
    \state{Illinois}
    \country{USA}
    }
    \email{feamster@uchicago.edu}

\author{Prateek Mittal}
\affiliation{%
    \institution{Princeton University} 
    \city{Princeton}
    \state{New Jersey}
    \country{USA}
    }
    \email{pmittal@princeton.edu}


\begin{document}
\fancyhead{}

\begin{sloppypar}
\begin{abstract}
    Machine learning is leveraged for many network traffic analysis tasks in security, from
    application identification to intrusion detection. Yet, the aspects of 
    the machine learning pipeline that ultimately determine the performance of the model---feature 
    selection and representation, model selection, and parameter tuning---remain manual 
    and painstaking. This paper presents a method to automate many aspects of traffic analysis,
    making it easier to apply machine learning techniques to a wider variety of
    traffic analysis tasks. 
    
    We introduce \name, a tool
    that generates a unified packet representation that is amenable for
    representation learning and model training. We integrate \name with
    automated machine learning (AutoML), resulting in \name{}ML, a public system 
    that largely eliminates feature extraction and model tuning for a wide variety of traffic analysis tasks.
    We have evaluated \name{}ML on eight separate traffic analysis tasks and released \name{} and \name{}ML
    to enable future work to extend these methods.
\end{abstract}

\maketitle
\section{Introduction}

Many traffic analysis tasks in network security rely on machine learning
(e.g., application
identification~\cite{karagiannis2005blinc,bernaille2006early,van2020flowprint}, device
identification~\cite{kohno2005remote, formby2016s}).  Although research has paid much
attention to the machine learning models applied to these tasks and the
performance of those models, in practice these tasks rely heavily on a
pipeline that involves manually engineering features and selecting and tuning
models. Arriving at an appropriate combination of features, model, and
model parameters is typically an iterative process.  Indeed, the effectiveness
of applying machine learning to network traffic analysis tasks often depends
on the selection and appropriate representation of features as much as the
model itself, yet this part of the process has remained exacting and manual.

Feature engineering and model selection are painstaking processes, typically
requiring substantial specialized domain knowledge to engineer features that
are both practical to measure or derive and result in an accurate model.  Even
with expert domain knowledge, feature exploration and engineering remains
largely a brittle and imperfect process, since the choice of features and how
to represent them can greatly affect model accuracy. Such manual extraction may
omit features that either were not
immediately apparent or involve complex relationships (e.g., non-linear
relationships between features). Furthermore, traffic patterns and conditions
invariably shift, rendering models and hand-crafted features
obsolete~\cite{gama2014survey,anderson2017machine}.
Finally, every new network detection or classification task requires re-inventing
the wheel: engineering new features, selecting appropriate models, and tuning new
parameters by hand. 

\begin{table*}[h]
\centering
 \resizebox{\textwidth}{!}{
\begin{tabular}{@{}lcr|crrrrcc@{}}
\toprule
\multicolumn{3}{c|}{\textbf{Problem Overview}} & \multicolumn{5}{c}{\textbf{nPrintML}} & \multicolumn{2}{c}{\textbf{Comparison}} \\ \midrule
\multicolumn{1}{c}{Description} & Dataset & \multicolumn{1}{c|}{\# Classes} &
    \multicolumn{1}{c}{\begin{tabular}[c]{@{}c@{}}Configuration \\
        eAppendix~\ref{appendix:config})\end{tabular}}  & \multicolumn{1}{c}{\begin{tabular}[c]{@{}c@{}}Sample Size\\ (\# Packets)\end{tabular}} & \multicolumn{1}{c}{\begin{tabular}[c]{@{}c@{}}Balanced \\ Accuracy\end{tabular}} & \multicolumn{1}{c}{\begin{tabular}[c]{@{}c@{}}ROC\\ AUC\end{tabular}} & \multicolumn{1}{c|}{\begin{tabular}[c]{@{}c@{}}Macro\\ F1\end{tabular}} & Score & Source \\ \midrule
        Active Device Fingerprinting (\S\ref{sec:active}) & Network Device
    Dataset~\cite{holland2020classifying} & 15 & -4 -t -i & 21 & 95.4 & 99.7 &
    \multicolumn{1}{r|}{95.5} & 92.9 (Macro-F1) &
    \begin{tabular}[c]{@{}c@{}}ML-Enhanced\\ Nmap~\cite{nmap}\end{tabular} \\ \midrule
        \multirow{4}{*}{Passive OS Detection (\S\ref{sec:passive}) } &
        \multirow{4}{*}{CICIDS 2017~\cite{sharafaldin2018toward}} &
        \multirow{3}{*}{3} & \multirow{4}{*}{-4 -t} & 1 & 99.5 & 99.9 &
        \multicolumn{1}{r|}{99.5} & \multirow{3}{*}{81.3 (Macro-F1)} &
        \multirow{4}{*}{p0f~\cite{p0f}} \\
 &  &  &  & 10 & 99.9 & 100 & \multicolumn{1}{r|}{99.9} &  &  \\
 &  &  &  & 100 & 99.9 & 100 & \multicolumn{1}{r|}{99.9} &  &  \\
 &  & 13 &  & 100 & 77.1 & 97.5 & \multicolumn{1}{r|}{76.9} & No Previous Work &  \\ \midrule
        \multirow{6}{*}{Application Identification via DTLS Handshakes (\S\ref{sec:app})} &
    \multirow{6}{*}{DTLS Handshakes~\cite{macmillan2020evaluating}} &
    \multirow{6}{*}{7} & -4 & \multirow{6}{*}{43} & 99.8 & 96.9 &
    \multicolumn{1}{r|}{99.7} & \multirow{6}{*}{99.8 (Average Accuracy)} &
    \multirow{6}{*}{Hand-Curated Features~\cite{macmillan2020evaluating}} \\
 &  &  & -u &  & 99.9 & 99.7 & \multicolumn{1}{r|}{99.5} &  &  \\
 &  &  & -p 10 &  & 95.0 & 78.8 & \multicolumn{1}{r|}{77.4} &  &  \\
 &  &  & -p 25 &  & 99.9 & 99.7 & \multicolumn{1}{r|}{99.7} &  &  \\
 &  &  & -p 100 &  & 99.9 & 99.7 & \multicolumn{1}{r|}{99.7} &  &  \\
 &  &  & -4 -u -p 10 &  & 99.8 & 99.9 & \multicolumn{1}{r|}{99.8} &  &  \\ \midrule
            \multirow{2}{*}{Malware Detection for IoT Traces (\S\ref{subsec:netml})} & \multirow{2}{*}{netML
    IoT ~\cite{barut2020netml,stratosphere}} & 2 &
    \multirow{2}{*}{-4 -t -u} & \multirow{2}{*}{10} & 92.4 & 99.5 &
    \multicolumn{1}{r|}{93.2} & 99.9 (True Positive Rate) &
    \multirow{8}{*}{\begin{tabular}[c]{@{}c@{}}NetML Challenge\\
    Leaderboard~\cite{netmlleaderboard}\end{tabular}} \\
 &  & 19 &  &  & 86.1 & 96.9 & \multicolumn{1}{r|}{84.1} & 39.7 (Balanced F1) &  \\ \cmidrule(r){1-9}
        \multirow{4}{*}{Type of Traffic in Capture (\S\ref{subsec:netml})} & \multirow{4}{*}{netML
    Non-VPN ~\cite{barut2020netml,draper2016characterization}} & \multirow{2}{*}{7} & -4 -t -u -p 10 & \multirow{4}{*}{10} & 81.9 & 98.0 & \multicolumn{1}{r|}{79.5} & \multirow{2}{*}{67.3 (Balanced F1)} &  \\
 &  &  & \multirow{3}{*}{-4 -t -u} &  & 76.1 & 94.2 & \multicolumn{1}{r|}{75.8} &  &  \\
 &  & 18 &  &  & 66.2 & 91.3 & \multicolumn{1}{r|}{63.7} & 42.1 (Balanced F1) &  \\
 &  & 31 &  &  & 60.9 & 92.2 & \multicolumn{1}{r|}{57.6} & 34.9 (Balanced F1) &  \\ \cmidrule(r){1-9}
        \multirow{2}{*}{Intrusion Detection (\S\ref{subsec:netml})} & \multirow{2}{*}{netML CICIDS
    2017~\cite{barut2020netml,sharafaldin2018toward}} & 2 & \multirow{2}{*}{-4 -t -u} & \multirow{2}{*}{5} & 99.9 & 99.9 & \multicolumn{1}{r|}{99.9} & 98.9 (True Positive Rate) &  \\
 &  & 8 &  &  & 99.9 & 99.9 & \multicolumn{1}{r|}{99.9} & 99.2 (Balanced F1) &  \\ \midrule
        \begin{tabular}[c]{@{}l@{}}Determine Country of Origin for \\ Android \& iOS Application Traces\end{tabular} (\S\ref{subsec:mobile}) & Cross Platform~\cite{crossmarketdataset} & 3 & -4 -t -u -p 50 & 25 & 96.8 & 90.2 & \multicolumn{1}{r|}{90.4} & \multicolumn{2}{c}{No Previous Work} \\ \midrule
        \multirow{3}{*}{\begin{tabular}[c]{@{}l@{}}Identify streaming video (DASH) \\ service via device SYN packets\end{tabular}(\S\ref{subsec:video})}  & \multirow{3}{*}{Streaming Video Providers~\cite{bronzino2019inferring}} & \multirow{3}{*}{4} & \multirow{3}{*}{-4 -t -u -R} & 10 & 77.9 & 96.0 & \multicolumn{1}{r|}{78.9} & \multicolumn{2}{c}{\multirow{3}{*}{No Previous Work}} \\
 &  &  &  & 25 & 90.2 & 98.6 & \multicolumn{1}{r|}{90.4} & \multicolumn{2}{c}{} \\
 &  &  &  & 50 & 98.4 & 99.9 & \multicolumn{1}{r|}{98.6} & \multicolumn{2}{c}{} \\ \bottomrule
\end{tabular}
}%
    \caption{Case studies we have performed with \name{}ML. \name{}ML enables
    automated high performance traffic analysis across a wide range and
    combination of protocols. In many cases, models trained on \name{} are
    able to outperform state-of-the-art tools with no hand-derived features.}
    \label{tab:case-studies}
\end{table*}

This paper reconsiders long-held norms in applying machine learning to network
traffic analysis; namely, we seek to reduce reliance upon human-driven feature
engineering. To do so, we explore whether and how a single, standard
representation of a network packet can serve as a building block for the
automation of many common traffic analysis tasks. Our goal is not to retread any
specific network classification problem, but rather to argue that many of these
problems can be made easier---and in some cases, completely automated---with a
unified representation of traffic that is amenable for input to existing
automated machine learning (AutoML) pipelines~\cite{erickson2020autogluon}.  To
demonstrate this capability, we designed a standard packet representation, {\em
\name{}}, that encodes each packet in an inherently normalized, binary
representation while preserving the underlying semantics of each packet. \name{}
enables machine learning models to automatically discover important features
from sets of packets for each distinct classification task without the need for
manual extraction of features from stateful, semantic network protocols. This
automation paves the way for faster iteration and deployment of machine learning
algorithms for networking, lowering the barriers to practical deployment.

The integration of \name{} with AutoML, a system we name \name{}ML, enables
automated model selection and hyperparameter tuning, enabling the creation of
complete traffic analysis pipelines with \name{}---often without writing a
single line of code. Our evaluation shows that models trained on \name{} can
perform fine-grained OS detection, achieve higher accuracy
than Nmap~\cite{nmap} in device fingerprinting, and automatically identify
applications in traffic with retransmissions. Further, we compare the
performance of models trained on \name{} to the best results from
netML, a public traffic analysis
challenge for malware detection, intrusion detection, and application
identification tasks~\cite{barut2020netml,netmlleaderboard}. \name{}ML outperforms the best performing
hand-tuned models trained on the extracted features in all cases but one.
Finally, to explore generality of \name{}ML for a broader range of
traffic analysis tasks, we use \name{}ML to train models on public datasets to
identify the country of origin for mobile application traces and to identify
streaming video service providers~\cite{bronzino2019inferring,crossmarketdataset}.

Table \ref{tab:case-studies} highlights the performance of
\name{}ML. To enable
others to replicate and extend these results---and to apply \name{} and
\name{}ML to a broader class of problems---we have publicly released \name{},
\name{}ML, and all datasets used in this work, to serve as a
benchmark for the research community and others.

Many problems, such as capturing temporal relationships across multiple traffic
flows, and running \name{}ML on longer traffic sequences remain unsolved and
unexplored. In this regard, we view \name{} and \name{}ML as the first
chapter in a new direction for research at the intersection of machine
learning and networking that enables researchers to focus less time on fine-tuning
models and features and more time interpreting and deploying the best models in
practice.

\begin{figure*}[t]
    \centering
    \includegraphics[scale=0.50]{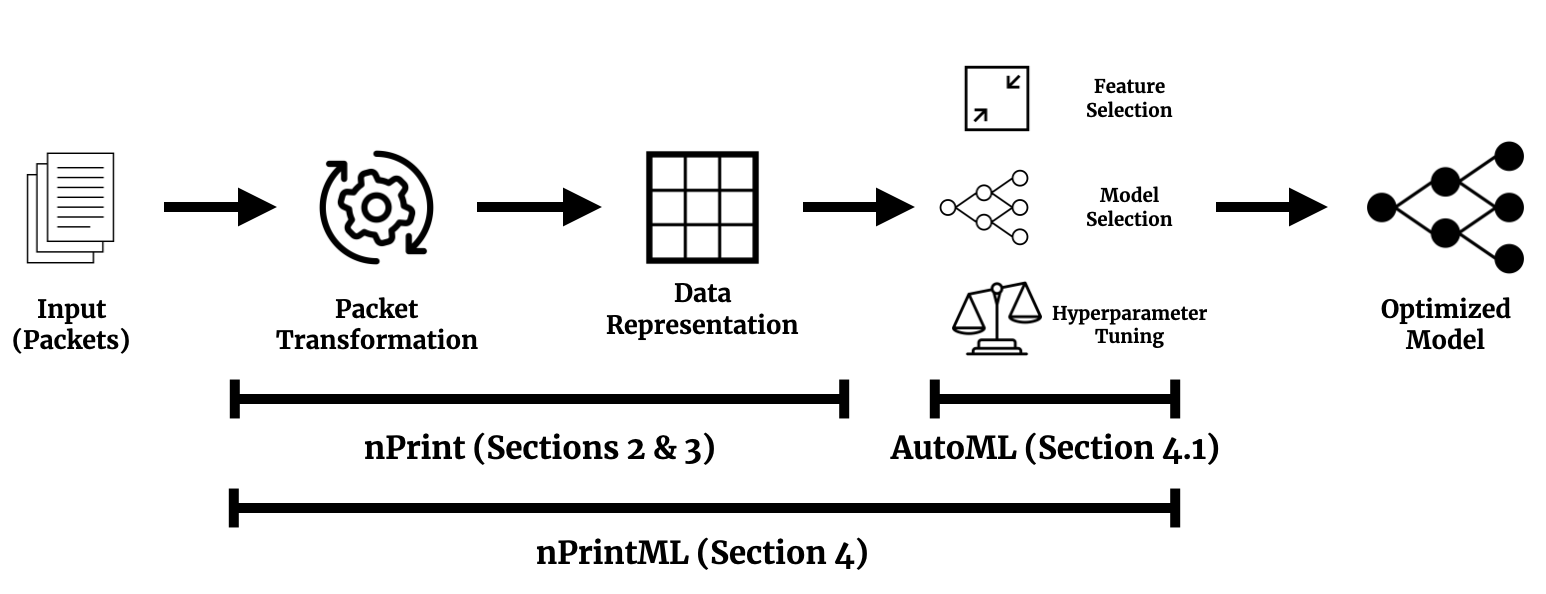}
    \caption{\name{} produces a standard network traffic representation that can
    be combined with AutoML tools to create \name{}ML, a standard, largely
    automated traffic analysis system.}
    \label{fig:system}
\end{figure*}

Figure~\ref{fig:system} summarizes the following sections of this paper.
Sections~\ref{sec:design} and ~\ref{sec:implement} discuss the design and
implementation of \name{}. Section~\ref{sec:nprintml} introduces
AutoML and \name{}ML, which combines AutoML and \name{}.
Section~\ref{sec:case-studies} applies \name{}ML for the
case studies in Table~\ref{tab:case-studies}. Finally,
Sections~\ref{sec:background} and~\ref{sec:conclusion} examine related works and
summarize our contributions.

\section{Data Representation}
\label{sec:design}

For many classification problems, 
the data representation is at least as important as the choice of model. Many machine learning models are well-tuned for standard
benchmarks (e.g., images, video, audio), but unfortunately network traffic
does not naturally lend itself to these types of representations. Nonetheless,
the nature of the models dictate certain design requirements, which we outline
in Section~\ref{sec:requirements}. Section~\ref{sec:representations} explores
three possible standard representations of network traffic, including a 
semantic encoding, an unaligned binary representation, and a hybrid approach, 
which fuses a binary representation of the packet with an encoding that simultaneously 
preserves the semantic structure of the bits.

\subsection{Design Requirements}\label{sec:requirements}

\paragraph{Complete.} Rather than select for certain features (or
representations), we aim to devise a representation that includes every bit of
a packet header. Doing so avoids the problems of relying on domain knowledge
that one packet header field (or combination of fields) is more important than others.
Our intuition is that the models can often determine which features are
important for a given problem without human guidance, given
a complete representation.

\paragraph{Constant size per problem.} Many machine learning models assume that inputs are
always the same size. For example, a trained neural network expects a certain
number of inputs; other models such as random forests and even regression
models expect that the input conforms to a particular size (i.e., number of
elements). Thus, each representation must be constant-size---even if the sizes
of individual packets or packet headers vary. Common approaches used in image
recognition, such as scaling, do not directly apply to network traffic. An ideal representation
has a size per-problem that is determined {\em a priori}; this 
knowledge avoids the need for multiple passes over a stored packet trace,
and it is essential in data streaming contexts.

\paragraph{Inherently normalized.} Machine learning models typically perform
better when features are normalized: even simple linear models (e.g., linear
and logistic regression) operate on normalized features; in more complex
models that use gradient-based optimization, normalization decreases
training time and increases model
stability~\cite{nayak2014impact,singh2019investigating}.  Of course, it is
possible to normalize {\em any} feature set, but it is more convenient if the
initial representation is inherently normalized.

\paragraph{Aligned.}
Every location in the representation should correspond to the same part of the
packet header across all packets. Alignment allows for models to learn
feature representations based on the fact that specific features (i.e.,
packet headers) are always located at the same offset in a packet. While human-driven feature engineering
leads to aligned features by extracting information from each packet into a well-formatted structure, this 
requirement is needed when considering packets in binary form, as both protocols and packets differ in length. 
Any misaligned features inject noise into the learning process, reducing the
accuracy
of the trained model.

\subsection{Building a Standard Data Representation}\label{sec:representations}

Network traffic can be represented in multiple ways. We discuss three
options: semantic, unaligned binary, and a hybrid
representation that we term \name{}.

\subsubsection{Semantic Representation} \label{sec:semantic}

\begin{figure*}[t]
    \centering
    \includegraphics[scale=0.70]{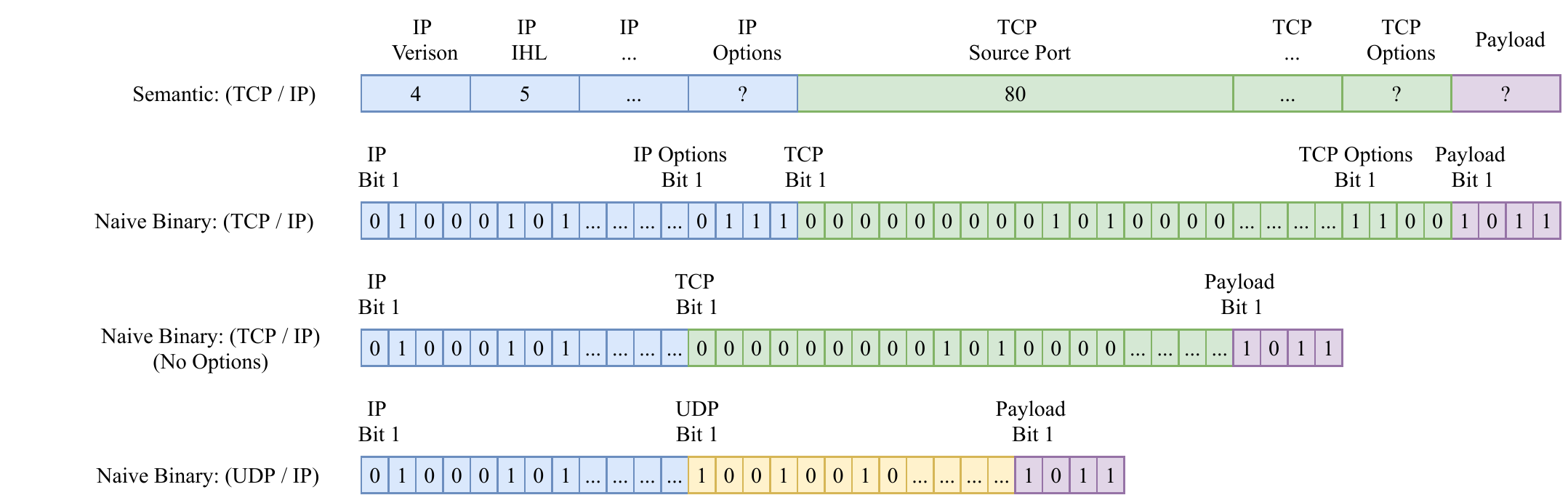}
    \caption{Semantic and naive binary \name{} representations introduce
        various problems that make modeling difficult: Semantic 
        loses ordering of specific fields such as the TCP
        options. Binary \name{}s lack alignment, which can
        degrade performance due to misaligned features.}
    \label{fig:bad-rep}
\end{figure*}

A classic view of network traffic examines packets as a collection of 
higher-level headers, such as IP, TCP, and UDP. Each header has 
semantic fields such as the IP TTL, the TCP port numbers, and the UDP
length fields. A standard semantic representation of network traffic collects
all of these semantic fields in a single representation. This semantic
representation is complete and constant size but has drawbacks.
Figure~\ref{fig:bad-rep} shows an example of this semantic representation and
some of its drawbacks as a standard representation. First, the representation does not preserve ordering of
options fields, which have been long used to separate classes of devices in
fingerprinting~\cite{nmap,smart2000defeating}.

We examine two of the datasets (presented in Sections~\ref{sec:passive} and
~\ref{sec:active}) and find 10 and 59 unique TCP option orderings, respectively. 
We further explore the effect that option ordering can have on classification 
performance in Appendix~\ref{appendix-ordering}, finding that a binary representation of
the TCP options, which preserves ordering, outperforms a semantic representation by ~10\% across a wide range of models 
in terms of F1-score. Second, domain expertise is
required to parse the semantic structure of every protocol, and even with this
knowledge, determining the correct representation of each feature is often a
significant exercise. For example, domain knowledge might indicate that the
TCP source port is an important field, but further (often manual) evaluation
may be needed to determine whether it should be represented as a continuous
value, or with a one-hot encoding, as well as if the feature needs to be normalized
before training. These decisions must be made for \textit{every} field extracted 
in a semantic manner, from IP addresses to each unique TCP option to ICMP address masks.

\subsubsection{Naive Binary Representation}

We can preserve ordering and mitigate reliance on manual feature engineering
with a raw bitmap representation. This choice leads to a consistent,
pre-normalized representation akin to an ``image'' of each packet. We
see an example of this representation in
Figure~\ref{fig:bad-rep}. However, transforming each packet into its bitmap
representation ignores many intricate details, including varying
sizes and protocols. These issues can cause feature vectors for two packets to have 
different meanings for the same feature. For example, a TCP packet and a UDP
packet with the same IP header would have entirely different information
represented as the same feature. Figure~\ref{fig:bad-rep} illustrates this
problem.

Such misalignment can occur between two packets of the same protocol: 
a TCP/IP packet with IP options and a TCP/IP packet without IP
options will cause the bits to be misaligned in the two representations.
Misalignment can manifest in two ways: 1) decreasing model performance as
the misaligned bits introduce noise in the model where important features may
exist; and 2) the resulting representation is
not interpretable, as it is impossible to map each bit in the
representation to a semantic meaning. The ability to understand the
features that determine the performance of a given model is especially
important in network traffic where 
the underlying data has semantic structure, in contrast to image classification,
where the underlying data is pixels.

\subsubsection{Hybrid: \name{}}

\begin{figure*}[t]
    \centering
    \includegraphics[scale=0.60]{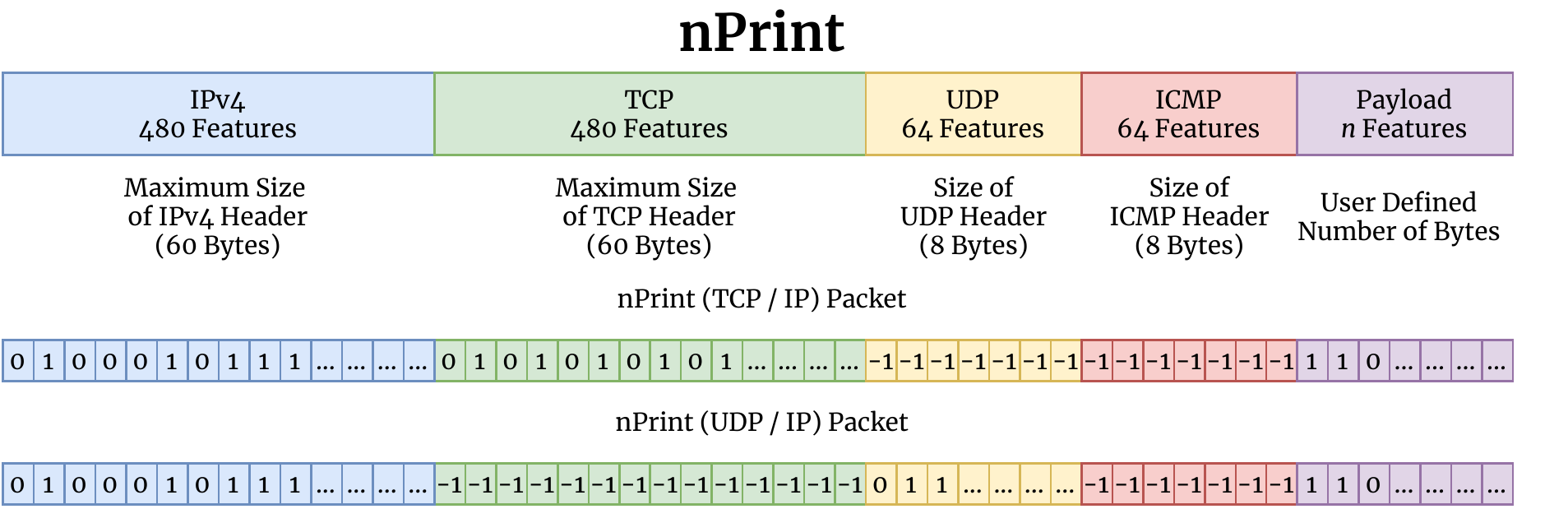}
    \caption{\name{}, the complete, normalized, aligned packet representation. Headers that do not exist in the
        packet being transformed are noted with sentinels accordingly, while headers that exist but are not of
        maximum size are zero padded for alignment across \name{}s. \name{} 
        removes reliance on manual feature engineering while
        avoiding misaligned features.}
    \label{fig:rep_diag}
\end{figure*}

Figure~\ref{fig:rep_diag} shows \name{}, a single-packet representation
that can be provided, unmodified, as input to machine learning models.
\name{} is a hybrid of semantic and binary packet representations,
representing packets to models as raw binary data, but aligning the binary data
in such a way that recognizes that the packets themselves have specific semantic
structure. By using internal padding, \name{}
mitigates misalignment that can occur with an unaligned binary representation while still preserving ordering of
options. Further, \name{} encodes the semantic structure of protocols while \textit{only} requiring knowledge of
the maximum length of the protocol compared to parsing and representing each field in a protocol.

\name{} is complete, aligned, constant size per problem, and
inherently normalized. \name{} is complete: any packet can be represented without
information loss. It is aligned: using internal padding and including space
for each header type regardless of whether that header is actually present in
a given packet ensures that each packet is represented in the same number of
features, and that each feature has the same meaning. Alignment gives \name{}
a distinct advantage over many network representations in that it is
interpretable at the bit level. This allows for researchers and practitioners
to map \name{} back to the semantic realm to better understand the features
that are driving the performance of a given model.  Not all models are
interpretable, but by having an interpretable \textit{representation}, we can
better understand models that are. \name{} is also inherently normalized: by directly
using the bits of the packets and filling non-existing headers with -1, each feature
takes on one of three values: -1, 0, or 1, removing the need for parsing and representing the values for each field in every packet. 
Further, filling non-existing header values with -1 allows us to differentiate between a bit being set to 0 and a header that
does not exist in the packet. Finally, \name{} is constant size per problem: each
packet is represented in the same number of features. We
make the payload an optional number of bytes for a given problem: with the
increasing majority of network traffic being encrypted, the payload is unusable
for many traffic classification problems.

\name{} is modular and extensible. First, other protocols (e.g., ICMP) can be 
added to the representation. Second, \name{}, which is a single-packet
representation, can be used as a building block for classification problems that
require sets of packets, as we have shown in several cases. 
If we consider each \name{} fingerprint as a 1xM matrix, where M is
the number of features in the fingerprint, we can build multi-packet \name{}
fingerprints by concatenating them.

\section{\name{} Implementation}
\label{sec:implement}

We implemented \name{} in C++ and evaluated it in different contexts,
including its memory footprint and a proof-of-concept evaluation on an operational 10 Gbps link.
\name{} currently supports 
Ethernet, IPv4, fixed IPv6 headers, UDP, TCP, ICMP, and any corresponding
packet payloads. Appendix~\ref{appendix:config} shows the full configuration
options available in \name{}. \name{} can either 
process offline packet packet capture formats such as PCAP and 
Zmap~\cite{zmap} output, or capture packets directly from a live interface. 
\name{} can also reverse the encoding, creating a PCAP from the \name format.
The \name{} source code is publicly available.

\paragraph{\name{} transforms over 1 million packets per minute.} We evaluate
the performance of \name{} on a system with a 4-core, 2.7~GHz CPU (Intel Core
i7-8559U) and 32GB of RAM. On this machine, \name{} transformed the three
datasets used in Section~\ref{sec:case-studies}, containing approximately 2.35
million, 274K, and 49K packets, in 49, 13, and 12 seconds
respectively\footnote{Each dataset is in a different file format resulting in
varying packet processing rates. The file formats are a single PCAP, a CSV of
hex-encoded raw packets, and over 7,000 PCAPs, respectively.} On average,
\name{} currently transforms packets at about 1.5 million packets per minute
using a single thread. We note here a few performance bottlenecks of the \name{}
implementation. \name{} writes output in a structured CSV format, chosen for
interoperability with popular machine learning libraries. Replacing \name{}'s
output to a binary format could yield significant performance benefits. Further,
\name{} leverages libpcap for packet processing. Leveraging optimized packet
libraries, such as zero-copy pf\_ring could also increase the performance of
\name{} on live traffic~\cite{pfring,libpcap}. We leave such optimizations for
future work.

\paragraph{\name{} has a constant memory footprint.}
\name{} has a constant memory footprint that depends only on the output
configuration. As an example, we profile memory usage with valgrind while
configuring \name{} to include IPv4 and TCP output (used for p0f evaluation in Section~\ref{sec:passive}). This configuration results
in a constant memory footprint of about 295~KB. A configuration consisting of  IPv4, UDP, ICMP, TCP, and 10
payload bytes yields a constant memory footprint of about 310~KB.

\paragraph{Proof of Concept: \name{} on an operational link.} \name{}
processes each packet independently, making it amenable to parallelization. We
run a proof-of-concept evaluation to understand of the baseline performance of
our public \name{} implementation on a 10~Gbps link comprised of traffic from a
consortium of universities using a commodity server. The server has two Intel
Xeon 6154 CPUs, each with 18 cores, running at 3~GHz and 376~GB of RAM. We
observe that by load balancing the traffic across multiple receive queue/CPU
pairs using Receive-Side Scaling (RSS\footnote{RSS is Intel-specific, depending
on hardware another similar multi-queue receive technology could also be used.})
we can run multiple, parallel \name{} processes without incurring penalties for
moving data between cores. We verified that \name{} is able to run on a live
traffic load of roughly 8 Gbps with near zero loss using 16 queues and 16
\name{} processes. Given this performance, \name{} should be capable of
processing higher rates by leveraging further parallelization and optimized
packet libraries, such as zero-copy pf\_ring~\cite{pfring}. We hope that our
publicly available implementation serves as groundwork for those who wish to
further optimize performance.

\section{\name{}ML}
\label{sec:nprintml}

Much previous work has used a pipeline similar to the full pipeline seen in
Figure~\ref{fig:system}. Packets are transformed into features developed by
experts over the course of days, weeks, or years, and ultimately trained on one
(or a small set of) models that experts believe will work best on the developed
features. Finally, the models are tuned either by hand or through a search process such as a grid search~\cite{sklearn:gs}. 
We highlight an opportunity to simplify the standard pipeline in
Figure~\ref{fig:system} through \name{} and new AutoML systems. 

First, we designed \name{} to be directly used in a machine learning pipeline,
standardizing the tedious feature development process for a large set of
traffic analysis problems. Next, our design choices for \name{} enable it to be
combined with new AutoML tools to standardize the second half of the
pipeline.

AutoML tools are designed to automate feature selection, model selection, and
hyperparameter tuning to find an optimized model for a given set of features and
labels. Rather than running hyperparameter optimization on one model, AutoML
tools use optimization techniques to perform combined model selection and
hyperparameter optimization, searching for the highest performing model in a
more principled manner than by hand.

We build upon our guiding principle that models can extract the best features for each
task and allow AutoML to determine the best type of model and hyperparameters
for that problem. This decision provides multiple benefits: 1) we can train and
test more model types, 2) we can optimize the hyperparameters for
\textit{every} model we train, and 3) we are certain that the best model is
chosen for a given representation. Finally, we note that although the case studies 
in this work and the public implementations of \name{} and \name{}ML focus on supervised
learning, there is opportunity for future work in combining the \name{} representation with
unsupervised learning techniques for other applications.

\subsection{AutoGluon-Tabular AutoML}

We use AutoGluon-Tabular to perform feature selection, model search, and
hyperparameter optimization for all eight problems we
evaluate~\cite{erickson2020autogluon}. We choose AutoGluon as it has been shown
to outperform many other public AutoML tools given the same
data~\cite{erickson2020autogluon}, and it is open source, though \name{}'s well-structure format 
make it amenable for any AutoML library. While many AutoML tools search a set of models and corresponding hyperparameters, AutoGluon
achieves higher performance by ensembling multiple single models that perform
well. AutoGluon-Tabular allows us to train, optimize, and test over 50 models
for each problem stemming from 6 different base model classes which are  
variations of tree-based methods, deep neural networks and neighbors-based classification. 
The highest performing model for each problem we examine is an ensemble of the base model classes.

AutoGluon has a \texttt{presets} parameter that determines the speed of the
training and size of the model versus the overall predictive quality of the models trained.
We set the \texttt{presets} parameter to
\verb|high_quality_fast_inference_only_refit|, which produces models with high predictive
accuracy and fast inference. There is a quality preset of ``best
quality'' which can create models with slightly higher predictive accuracy, but
at the cost of \textasciitilde10x-200x slower inference and
\textasciitilde10x-200x higher disk usage. We make this decision as we believe
inference time is an important metric when considering network traffic analysis.
By selecting a high quality model setting, each model is bagged with 10 folds, decreasing the
bias of individual models. We note that the preset parameter for an AutoML tool does not represent the training of 
a single model, but an optimization of a set of models for a given task.

We set no limit on model training time, allowing AutoGluon to find the
best model, and split every dataset into 75\%
training and 25\% testing datasets. Finally, we set the evaluation metric to
\verb|f1_macro|, which represents a F1 score that is calculated by calculating
the F1 Score for each class in a multi-class classification problem and
calculating their unweighted mean. This decision leads AutoGluon to tune hyperparameters and ensemble
weights to optimize the F1-macro score on validation data.

\subsection{\name{}ML Implementation}

We have implemented \name{}ML in Python, directly combining \name{} and
AutoGluon-Tabular AutoML. \name{}ML enables researchers to create full traffic
analysis pipelines using \name{} in a single program call. Below is an example
of fully recreating the passive OS detection case study found in
Section~\ref{sec:case-studies}.

\begin{lstlisting}
$ nprintml -L os_labels.txt -a index \
-P traffic.pcap -4 -t --sample_size 10
\end{lstlisting}

A full tutorial of \name{}ML is supplied in Appendix~\ref{appendix-tutorial}.
At the time of writing, \name{}ML can create machine learning pipelines from a
single traffic trace or an entire directory of traffic traces.

\paragraph{Metrics.}
\name{}ML outputs a standard set of metrics for each classification problem,
which we define here. We define a \textit{false positive} for class \textit{C} as any sample that is
not of class \textit{C}, but misclassified as class \textit{C} by the
classifier. A \textit{false negative} for class \textit{C} is any sample of
class \textit{C} that is not classified as class \textit{C}. We then evaluate
each trained model using multiple metrics including balanced accuracy, ROC AUC,
and F1 scores. We use a \textit{balanced accuracy score} to account for any
class imbalance in the data. In the multi-class classification case we present
macro AUC ROC scores in a ``one vs rest'' manner, where each class \textit{C} is
considered as a binary classification task between \textit{C} and each other
class. The ROC AUC is computed by calculating the ROC AUC for each class
and taking their unweighted mean. F1 scores represent a weighted average of
precision and recall. For multi-class classification, we report a macro F1
score, calculated in the same manner as optimized during training.

\section{Related Work}
\label{sec:background}

This section explores past work on manual and automated fingerprinting
techniques and
how they relate to \name.

\paragraph{Machine learning-based traffic analysis.}
Machine learning techniques have been applied to network traffic classification
and fingerprinting~\cite{alsabah,6104509,10.1145/1163593.1163596, venkataraman}.
Wang \etal{} developed an ML-based classification model to detect obfuscated
traffic~\cite{10.1145/2810103.2813715}. Sommer and Paxson demonstrated that
using machine learning to detect anomalies can have significant drawbacks, as
network anomalies can exhibit different behavior than other problems solved by
ML~\cite{sommer2010outside}. 
Trimananda \etal{} used DBSCAN to identify smart home device actions in network traffic~\cite{trimananda2020packet}.

Other work has used machine learning to
identify websites visited through the Tor network~\cite{197185, 184463,
panchenko2016website, 10.5555/3241189.3241296}.
Deep learning techniques have recently garnered attention as they have proven
to be applicable to the task for inferring information from encrypted network
traffic. Various work has used machine learning models to fingerprint websites
visited through the Tor
network~\cite{DBLP:conf/ndss/RimmerPJGJ18,sirinam2018deep,oh2019p1,10.1145/3243734.3243824}. These works
 differ from this work, due to their focus on the Tor setting. In
Tor, all packets are the same size and encrypted, meaning network traffic in Tor can be
represented by a series of -1s and 1s that represent the direction of the
traffic. This work in contrast considers traffic over any network that can vary in
size and protocol. 

Deep learning techniques have become popular for network traffic
classification problems~\cite{8004872,yu2017network,8640262,hwang2019lstm,kitsune}. Yu
\etal{} used convolutional autoencoders for network intrusion
detection~\cite{yu2017network}. Mirksy \etal{} use an ensemble of autoencoders and human-engineered features 
on the intrusion detection problem in an unsupervised setting~\cite{kitsune}. Our
work differs as we aim to explore whether a variety of machine learning models can automatically extract important features from network traffic
in a supervised setting. Wang \etal{} applied off-the-shelf deep learning techniques from image recognition and text analysis to intrusion detection; in
contrast, we focus specifically on creating a general representation for
network traffic that can be used in a variety of different models, across a
broad class of problems\cite{wang2017hast}.  Our results also suggest that Wang \etal{}'s model
could be more complex than necessary, and that better input representations such
as \name{} may result in simpler models.

\paragraph{TCP-based host fingerprinting.}
Idiosyncrasies between TCP/IP stack implementations have often been the basis of
networked host fingerprinting techniques. Actively probing to differentiate
between TCP implementations was introduced by Comer and
Lin~\cite{comer1994probing}. Padhye and Floyd identified differences between TCP
implementations to detect bugs in public web servers~\cite{10.1145/964723.383083}. Paxson passively identified TCP
implementations using traffic traces~\cite{10.1145/263109.263160}.

Past work has developed techniques to fingerprint host operating systems and
devices.
There are multiple tools and methods for host OS fingerprinting, using both
active and passive techniques. Passive OS identification aims to identify operating systems from passively
captured network traffic~\cite{robustclass,
lippmann2003passive}. P0f passively observes traffic and determines
the operating system largely based on TCP behavior~\cite{p0f}. 
Another common tool is Nmap~\cite{nmap}, which performs active fingerprinting.
Nmap sends probes and examines the responses received from a target host,
focusing on TCP/IP settings and Internet Control Messaging Protocol (ICMP)
implementation differences between different operating systems and devices.
Nmap is widely considered the ``gold standard'' of active probing tools.  In
contrast, \name does not focus on heuristics and \textit{a priori} knowledge
of implementation differences between host networking stacks. Instead, \name
relies on the model to learn these differences during training.

Remote fingerprinting can be used to characterize aspects of the remote system
other than its operating system or networking stack. Clock skew information
determined from the TCP timestamp option was used to identify individual
physical devices by Kohno \etal{}~\cite{kohno2005remote}. Formby \etal{}
passively fingerprint industrial control system devices~\cite{FormbySLRB16}.

\paragraph{Automated machine learning.}
While the use of deep learning techniques can help automate feature engineering, 
a separate line of research has examined how to perform automated machine
learning (AutoML). The work examines the use of optimization
techniques to automate not only feature engineering and selection, but
model archietecture and hyperparameter optimization as 
well~\cite{kotthoff2017auto,feurer2019auto,jin2019auto,ledell2020h2o,feurer2015efficient}. These
tools have recently been used for model compression, image classification, and even bank
failure prediction~\cite{agrapetidou2020automl,he2018amc}. To our knowledge, we are the
first to explore the combination of AutoML and network traffic classification.

We specifically use AutoGluon-Tabular, which performs feature selection,
model selection, and hyperparameter optimization by searching through a set of 
base models~\cite{erickson2020autogluon}. These models include deep neural networks, tree based methods such as
random forests, non-parametric methods such as K-nearest neighbors, and gradient
boosted tree methods. Beyond searching the singular models, AutoGluon-Tabular
creates weighted ensemble models out of the base models to achieve higher
performance than other AutoML tools in less overall training
time~\cite{erickson2020autogluon}.

\balance\section{Conclusion}\label{sec:conclusion}

The effectiveness of applying machine learning to traffic analysis
tasks in network security depends on the selection and appropriate representation of features as
much as, if not more than, the model itself. This paper creates a new
direction for automated traffic analysis, presenting \name{}, a unified packet
representation that takes as input raw network packets and transforms them
into a format that is amenable to representation learning and model training,
thereby automating a part of the machine learning process which until now has
been largely painstaking and manual.  

This standard format makes it easy to
integrate network traffic analysis with state-of-the-art automated machine learning
(AutoML) pipelines. \name{}ML, the integration of \name{} with AutoML,
automatically learns the best model, parameter settings, and feature
representation for the corresponding task. We applied \name{}ML to eight
common network traffic analysis tasks, improving on the state of the art in
many cases.  \name{} has demonstrated that many network traffic classification
tasks are amenable to automation, though many open problems exist such as
automated timeseries analysis and classification involving multiple flows.
\name{} should ultimately be applied to a larger set of classification
problems.  To this end, we have released \name{}, \name{}ML, and all datasets 
as a benchmark for the research community, to make it easy to both replicate
and extend the results from this work.

\section{Acknowledgements}
We thank our shepherd Katharina Kohls and the anonymous reviewers for their
helpful comments. We also thank Vitaly Shmatikov for guidance and feedback on early versions of this work,
and Jesse London for collaborating on the development of \name{}ML. This research was supported in part by the Center for
Information Technology Policy at Princeton University and by NSF Award
CPS-1739809 under a cooperative agreement with the Department of Homeland
Security, DARPA and AFRL under Contract FA8750-19-C-0079, and NSF Awards
CNS-1553437 and CNS-1704105.

\end{sloppypar}

\newpage
\bibliographystyle{ACM-Reference-Format}
\bibliography{proceedings,paper}


\begin{thebibliography}{64}


\ifx \showCODEN    \undefined \def \showCODEN     #1{\unskip}     \fi
\ifx \showDOI      \undefined \def \showDOI       #1{#1}\fi
\ifx \showISBNx    \undefined \def \showISBNx     #1{\unskip}     \fi
\ifx \showISBNxiii \undefined \def \showISBNxiii  #1{\unskip}     \fi
\ifx \showISSN     \undefined \def \showISSN      #1{\unskip}     \fi
\ifx \showLCCN     \undefined \def \showLCCN      #1{\unskip}     \fi
\ifx \shownote     \undefined \def \shownote      #1{#1}          \fi
\ifx \showarticletitle \undefined \def \showarticletitle #1{#1}   \fi
\ifx \showURL      \undefined \def \showURL       {\relax}        \fi
\providecommand\bibfield[2]{#2}
\providecommand\bibinfo[2]{#2}
\providecommand\natexlab[1]{#1}
\providecommand\showeprint[2][]{arXiv:#2}

\bibitem[\protect\citeauthoryear{{Aceto}, {Ciuonzo}, {Montieri}, and
  {Pescapé}}{{Aceto} et~al\mbox{.}}{2019}]%
        {8640262}
\bibfield{author}{\bibinfo{person}{G. {Aceto}}, \bibinfo{person}{D. {Ciuonzo}},
  \bibinfo{person}{A. {Montieri}}, {and} \bibinfo{person}{A. {Pescapé}}.}
  \bibinfo{year}{2019}\natexlab{}.
\newblock \showarticletitle{Mobile Encrypted Traffic Classification Using Deep
  Learning: Experimental Evaluation, Lessons Learned, and Challenges}.
\newblock \bibinfo{journal}{\emph{IEEE Transactions on Network and Service
  Management}} \bibinfo{volume}{16}, \bibinfo{number}{2}
  (\bibinfo{year}{2019}), \bibinfo{pages}{445--458}.
\newblock


\bibitem[\protect\citeauthoryear{Agrapetidou, Charonyktakis, Gogas,
  Papadimitriou, and Tsamardinos}{Agrapetidou et~al\mbox{.}}{2020}]%
        {agrapetidou2020automl}
\bibfield{author}{\bibinfo{person}{Anna Agrapetidou}, \bibinfo{person}{Paulos
  Charonyktakis}, \bibinfo{person}{Periklis Gogas}, \bibinfo{person}{Theophilos
  Papadimitriou}, {and} \bibinfo{person}{Ioannis Tsamardinos}.}
  \bibinfo{year}{2020}\natexlab{}.
\newblock \showarticletitle{An AutoML application to forecasting bank
  failures}.
\newblock \bibinfo{journal}{\emph{Applied Economics Letters}}
  (\bibinfo{year}{2020}), \bibinfo{pages}{1--5}.
\newblock


\bibitem[\protect\citeauthoryear{AlSabah, Bauer, and Goldberg}{AlSabah
  et~al\mbox{.}}{2012}]%
        {alsabah}
\bibfield{author}{\bibinfo{person}{Mashael AlSabah}, \bibinfo{person}{Kevin
  Bauer}, {and} \bibinfo{person}{Ian Goldberg}.}
  \bibinfo{year}{2012}\natexlab{}.
\newblock \showarticletitle{Enhancing {Tor’s} Performance Using Real-Time
  Traffic Classification}. In \bibinfo{booktitle}{\emph{Proceedings of the 2012
  ACM Conference on Computer and Communications Security}} (Raleigh, North
  Carolina, USA) \emph{(\bibinfo{series}{CCS ’12})}.
  \bibinfo{publisher}{Association for Computing Machinery},
  \bibinfo{address}{New York, NY, USA}, \bibinfo{pages}{73–84}.
\newblock
\showISBNx{9781450316514}
\urldef\tempurl%
\url{https://doi.org/10.1145/2382196.2382208}
\showDOI{\tempurl}


\bibitem[\protect\citeauthoryear{Anderson and McGrew}{Anderson and
  McGrew}{2017}]%
        {anderson2017machine}
\bibfield{author}{\bibinfo{person}{Blake Anderson} {and} \bibinfo{person}{David
  McGrew}.} \bibinfo{year}{2017}\natexlab{}.
\newblock \showarticletitle{{Machine Learning for Encrypted Malware Traffic
  Classification: Accounting for Noisy Labels and Non-stationarity}}. In
  \bibinfo{booktitle}{\emph{Proceedings of the 23rd ACM SIGKDD International
  Conference on knowledge discovery and data mining}}.
  \bibinfo{pages}{1723--1732}.
\newblock


\bibitem[\protect\citeauthoryear{{Barker}, {Hannay}, and {Szewczyk}}{{Barker}
  et~al\mbox{.}}{2011}]%
        {6104509}
\bibfield{author}{\bibinfo{person}{J. {Barker}}, \bibinfo{person}{P. {Hannay}},
  {and} \bibinfo{person}{P. {Szewczyk}}.} \bibinfo{year}{2011}\natexlab{}.
\newblock \showarticletitle{Using Traffic Analysis to Identify the Second
  Generation Onion Router}. In \bibinfo{booktitle}{\emph{2011 IFIP 9th
  International Conference on Embedded and Ubiquitous Computing}}.
  \bibinfo{pages}{72--78}.
\newblock
\showISSN{null}
\urldef\tempurl%
\url{https://doi.org/10.1109/EUC.2011.76}
\showDOI{\tempurl}


\bibitem[\protect\citeauthoryear{Barut, Luo, Zhang, Li, and Li}{Barut
  et~al\mbox{.}}{2020}]%
        {barut2020netml}
\bibfield{author}{\bibinfo{person}{Onur Barut}, \bibinfo{person}{Yan Luo},
  \bibinfo{person}{Tong Zhang}, \bibinfo{person}{Weigang Li}, {and}
  \bibinfo{person}{Peilong Li}.} \bibinfo{year}{2020}\natexlab{}.
\newblock \showarticletitle{NetML: A Challenge for Network Traffic Analytics}.
\newblock \bibinfo{journal}{\emph{arXiv preprint arXiv:2004.13006}}
  (\bibinfo{year}{2020}).
\newblock


\bibitem[\protect\citeauthoryear{Bellovin}{Bellovin}{2002}]%
        {10.1145/637201.637243}
\bibfield{author}{\bibinfo{person}{Steven~M. Bellovin}.}
  \bibinfo{year}{2002}\natexlab{}.
\newblock \showarticletitle{A Technique for Counting Natted Hosts}. In
  \bibinfo{booktitle}{\emph{Proceedings of the 2nd ACM SIGCOMM Workshop on
  Internet Measurment}} (Marseille, France) \emph{(\bibinfo{series}{IMW
  ’02})}. \bibinfo{publisher}{Association for Computing Machinery},
  \bibinfo{address}{New York, NY, USA}, \bibinfo{pages}{267–272}.
\newblock
\showISBNx{158113603X}
\urldef\tempurl%
\url{https://doi.org/10.1145/637201.637243}
\showDOI{\tempurl}


\bibitem[\protect\citeauthoryear{Bernaille, Teixeira, and Salamatian}{Bernaille
  et~al\mbox{.}}{2006}]%
        {bernaille2006early}
\bibfield{author}{\bibinfo{person}{Laurent Bernaille}, \bibinfo{person}{Renata
  Teixeira}, {and} \bibinfo{person}{Kave Salamatian}.}
  \bibinfo{year}{2006}\natexlab{}.
\newblock \showarticletitle{{Early Application Identification}}. In
  \bibinfo{booktitle}{\emph{Proceedings of the 2006 ACM CoNEXT Conference}}.
  \bibinfo{pages}{1--12}.
\newblock


\bibitem[\protect\citeauthoryear{Beverly}{Beverly}{2004}]%
        {robustclass}
\bibfield{author}{\bibinfo{person}{Robert Beverly}.}
  \bibinfo{year}{2004}\natexlab{}.
\newblock \showarticletitle{A Robust Classifier for Passive {TCP/IP}
  Fingerprinting}. In \bibinfo{booktitle}{\emph{Proceedings of the 5th Passive
  and Active Measurement ({PAM}) Workshop}}.
\newblock


\bibitem[\protect\citeauthoryear{Bronzino, Schmitt, Ayoubi, Martins, Teixeira,
  and Feamster}{Bronzino et~al\mbox{.}}{2019}]%
        {bronzino2019inferring}
\bibfield{author}{\bibinfo{person}{Francesco Bronzino}, \bibinfo{person}{Paul
  Schmitt}, \bibinfo{person}{Sara Ayoubi}, \bibinfo{person}{Guilherme Martins},
  \bibinfo{person}{Renata Teixeira}, {and} \bibinfo{person}{Nick Feamster}.}
  \bibinfo{year}{2019}\natexlab{}.
\newblock \showarticletitle{Inferring streaming video quality from encrypted
  traffic: Practical models and deployment experience}.
\newblock \bibinfo{journal}{\emph{Proceedings of the ACM on Measurement and
  Analysis of Computing Systems}} \bibinfo{volume}{3}, \bibinfo{number}{3}
  (\bibinfo{year}{2019}), \bibinfo{pages}{1--25}.
\newblock


\bibitem[\protect\citeauthoryear{Comer and Lin}{Comer and Lin}{1994}]%
        {comer1994probing}
\bibfield{author}{\bibinfo{person}{Douglas~E Comer} {and}
  \bibinfo{person}{John~C Lin}.} \bibinfo{year}{1994}\natexlab{}.
\newblock \showarticletitle{Probing {TCP} implementations}. In
  \bibinfo{booktitle}{\emph{Usenix Summer}}. \bibinfo{pages}{245--255}.
\newblock


\bibitem[\protect\citeauthoryear{Draper-Gil, Lashkari, Mamun, and
  Ghorbani}{Draper-Gil et~al\mbox{.}}{2016}]%
        {draper2016characterization}
\bibfield{author}{\bibinfo{person}{Gerard Draper-Gil},
  \bibinfo{person}{Arash~Habibi Lashkari}, \bibinfo{person}{Mohammad
  Saiful~Islam Mamun}, {and} \bibinfo{person}{Ali~A Ghorbani}.}
  \bibinfo{year}{2016}\natexlab{}.
\newblock \showarticletitle{Characterization of encrypted and vpn traffic using
  time-related}. In \bibinfo{booktitle}{\emph{Proceedings of the 2nd
  international conference on information systems security and privacy
  (ICISSP)}}. \bibinfo{pages}{407--414}.
\newblock


\bibitem[\protect\citeauthoryear{Durumeric, Wustrow, and Halderman}{Durumeric
  et~al\mbox{.}}{2013}]%
        {zmap}
\bibfield{author}{\bibinfo{person}{Zakir Durumeric}, \bibinfo{person}{Eric
  Wustrow}, {and} \bibinfo{person}{J~Alex Halderman}.}
  \bibinfo{year}{2013}\natexlab{}.
\newblock \showarticletitle{ZMap: Fast Internet-wide scanning and its security
  applications}. In \bibinfo{booktitle}{\emph{Proceeedings of the 22nd {USENIX}
  Security Symposium}}. \bibinfo{pages}{605--620}.
\newblock


\bibitem[\protect\citeauthoryear{Erickson, Mueller, Shirkov, Zhang, Larroy, Li,
  and Smola}{Erickson et~al\mbox{.}}{2020}]%
        {erickson2020autogluon}
\bibfield{author}{\bibinfo{person}{Nick Erickson}, \bibinfo{person}{Jonas
  Mueller}, \bibinfo{person}{Alexander Shirkov}, \bibinfo{person}{Hang Zhang},
  \bibinfo{person}{Pedro Larroy}, \bibinfo{person}{Mu Li}, {and}
  \bibinfo{person}{Alexander Smola}.} \bibinfo{year}{2020}\natexlab{}.
\newblock \showarticletitle{AutoGluon-Tabular: Robust and Accurate AutoML for
  Structured Data}.
\newblock \bibinfo{journal}{\emph{arXiv preprint arXiv:2003.06505}}
  (\bibinfo{year}{2020}).
\newblock


\bibitem[\protect\citeauthoryear{Feurer, Klein, Eggensperger, Springenberg,
  Blum, and Hutter}{Feurer et~al\mbox{.}}{2015}]%
        {feurer2015efficient}
\bibfield{author}{\bibinfo{person}{Matthias Feurer}, \bibinfo{person}{Aaron
  Klein}, \bibinfo{person}{Katharina Eggensperger}, \bibinfo{person}{Jost
  Springenberg}, \bibinfo{person}{Manuel Blum}, {and} \bibinfo{person}{Frank
  Hutter}.} \bibinfo{year}{2015}\natexlab{}.
\newblock \showarticletitle{Efficient and robust automated machine learning}.
  In \bibinfo{booktitle}{\emph{Advances in neural information processing
  systems}}. \bibinfo{pages}{2962--2970}.
\newblock


\bibitem[\protect\citeauthoryear{Feurer, Klein, Eggensperger, Springenberg,
  Blum, and Hutter}{Feurer et~al\mbox{.}}{2019}]%
        {feurer2019auto}
\bibfield{author}{\bibinfo{person}{Matthias Feurer}, \bibinfo{person}{Aaron
  Klein}, \bibinfo{person}{Katharina Eggensperger},
  \bibinfo{person}{Jost~Tobias Springenberg}, \bibinfo{person}{Manuel Blum},
  {and} \bibinfo{person}{Frank Hutter}.} \bibinfo{year}{2019}\natexlab{}.
\newblock \showarticletitle{Auto-sklearn: efficient and robust automated
  machine learning}.
\newblock In \bibinfo{booktitle}{\emph{Automated Machine Learning}}.
  \bibinfo{publisher}{Springer, Cham}, \bibinfo{pages}{113--134}.
\newblock


\bibitem[\protect\citeauthoryear{Formby, Srinivasan, Leonard, Rogers, and
  Beyah}{Formby et~al\mbox{.}}{2016a}]%
        {formby2016s}
\bibfield{author}{\bibinfo{person}{David Formby}, \bibinfo{person}{Preethi
  Srinivasan}, \bibinfo{person}{Andrew~M Leonard}, \bibinfo{person}{Jonathan~D
  Rogers}, {and} \bibinfo{person}{Raheem~A Beyah}.}
  \bibinfo{year}{2016}\natexlab{a}.
\newblock \showarticletitle{{Who's in Control of Your Control System? Device
  Fingerprinting for Cyber-Physical Systems}}. In
  \bibinfo{booktitle}{\emph{Network and Distributed Systems Security
  Symposium}}.
\newblock


\bibitem[\protect\citeauthoryear{Formby, Srinivasan, Leonard, Rogers, and
  Beyah}{Formby et~al\mbox{.}}{2016b}]%
        {FormbySLRB16}
\bibfield{author}{\bibinfo{person}{David Formby}, \bibinfo{person}{Preethi
  Srinivasan}, \bibinfo{person}{Andrew~M. Leonard},
  \bibinfo{person}{Jonathan~D. Rogers}, {and} \bibinfo{person}{Raheem~A.
  Beyah}.} \bibinfo{year}{2016}\natexlab{b}.
\newblock \showarticletitle{Who's in Control of Your Control System? Device
  Fingerprinting for Cyber-Physical Systems}. In \bibinfo{booktitle}{\emph{23rd
  Annual Network and Distributed System Security Symposium, {NDSS}, 2016, San
  Diego, California, USA, February 21-24, 2016}}.
\newblock
\urldef\tempurl%
\url{http://wp.internetsociety.org/ndss/wp-content/uploads/sites/25/2017/09/who-control-your-control-system-device-fingerprinting-cyber-physical-systems.pdf}
\showURL{%
\tempurl}


\bibitem[\protect\citeauthoryear{Gama, {\v{Z}}liobait{\.e}, Bifet, Pechenizkiy,
  and Bouchachia}{Gama et~al\mbox{.}}{2014}]%
        {gama2014survey}
\bibfield{author}{\bibinfo{person}{Jo{\~a}o Gama}, \bibinfo{person}{Indr{\.e}
  {\v{Z}}liobait{\.e}}, \bibinfo{person}{Albert Bifet}, \bibinfo{person}{Mykola
  Pechenizkiy}, {and} \bibinfo{person}{Abdelhamid Bouchachia}.}
  \bibinfo{year}{2014}\natexlab{}.
\newblock \showarticletitle{{A Survey on Concept Drift Adaptation}}.
\newblock \bibinfo{journal}{\emph{ACM Computing Surveys (CSUR)}}
  \bibinfo{volume}{46}, \bibinfo{number}{4} (\bibinfo{year}{2014}),
  \bibinfo{pages}{1--37}.
\newblock


\bibitem[\protect\citeauthoryear{Hayes and Danezis}{Hayes and Danezis}{2016}]%
        {197185}
\bibfield{author}{\bibinfo{person}{Jamie Hayes} {and} \bibinfo{person}{George
  Danezis}.} \bibinfo{year}{2016}\natexlab{}.
\newblock \showarticletitle{k-fingerprinting: A Robust Scalable Website
  Fingerprinting Technique}. In \bibinfo{booktitle}{\emph{25th {USENIX}
  Security Symposium ({USENIX} Security 16)}}. \bibinfo{publisher}{{USENIX}
  Association}, \bibinfo{address}{Austin, TX}, \bibinfo{pages}{1187--1203}.
\newblock
\showISBNx{978-1-931971-32-4}
\urldef\tempurl%
\url{https://www.usenix.org/conference/usenixsecurity16/technical-sessions/presentation/hayes}
\showURL{%
\tempurl}


\bibitem[\protect\citeauthoryear{He, Lin, Liu, Wang, Li, and Han}{He
  et~al\mbox{.}}{2018}]%
        {he2018amc}
\bibfield{author}{\bibinfo{person}{Yihui He}, \bibinfo{person}{Ji Lin},
  \bibinfo{person}{Zhijian Liu}, \bibinfo{person}{Hanrui Wang},
  \bibinfo{person}{Li-Jia Li}, {and} \bibinfo{person}{Song Han}.}
  \bibinfo{year}{2018}\natexlab{}.
\newblock \showarticletitle{Amc: Automl for model compression and acceleration
  on mobile devices}. In \bibinfo{booktitle}{\emph{Proceedings of the European
  Conference on Computer Vision (ECCV)}}. \bibinfo{pages}{784--800}.
\newblock


\bibitem[\protect\citeauthoryear{Holland, Teixeria, Schmitt, Borgolte, Rexford,
  Feamster, and Mayer}{Holland et~al\mbox{.}}{2020}]%
        {holland2020classifying}
\bibfield{author}{\bibinfo{person}{Jordan Holland}, \bibinfo{person}{Ross
  Teixeria}, \bibinfo{person}{Paul Schmitt}, \bibinfo{person}{Kevin Borgolte},
  \bibinfo{person}{Jennifer Rexford}, \bibinfo{person}{Nick Feamster}, {and}
  \bibinfo{person}{Jonathan Mayer}.} \bibinfo{year}{2020}\natexlab{}.
\newblock \showarticletitle{Classifying Network Vendors at Internet scale}.
\newblock \bibinfo{journal}{\emph{arXiv preprint arXiv:2006.13086}}
  (\bibinfo{year}{2020}).
\newblock


\bibitem[\protect\citeauthoryear{Hwang, Peng, Nguyen, and Chang}{Hwang
  et~al\mbox{.}}{2019}]%
        {hwang2019lstm}
\bibfield{author}{\bibinfo{person}{Ren-Hung Hwang}, \bibinfo{person}{Min-Chun
  Peng}, \bibinfo{person}{Van-Linh Nguyen}, {and} \bibinfo{person}{Yu-Lun
  Chang}.} \bibinfo{year}{2019}\natexlab{}.
\newblock \showarticletitle{An LSTM-Based Deep Learning Approach for
  Classifying Malicious Traffic at the Packet Level}.
\newblock \bibinfo{journal}{\emph{Applied Sciences}} \bibinfo{volume}{9},
  \bibinfo{number}{16} (\bibinfo{year}{2019}), \bibinfo{pages}{3414}.
\newblock


\bibitem[\protect\citeauthoryear{Jin, Song, and Hu}{Jin et~al\mbox{.}}{2019}]%
        {jin2019auto}
\bibfield{author}{\bibinfo{person}{Haifeng Jin}, \bibinfo{person}{Qingquan
  Song}, {and} \bibinfo{person}{Xia Hu}.} \bibinfo{year}{2019}\natexlab{}.
\newblock \showarticletitle{Auto-keras: An efficient neural architecture search
  system}. In \bibinfo{booktitle}{\emph{Proceedings of the 25th ACM SIGKDD
  International Conference on Knowledge Discovery \& Data Mining}}.
  \bibinfo{pages}{1946--1956}.
\newblock


\bibitem[\protect\citeauthoryear{Karagiannis, Papagiannaki, and
  Faloutsos}{Karagiannis et~al\mbox{.}}{2005}]%
        {karagiannis2005blinc}
\bibfield{author}{\bibinfo{person}{Thomas Karagiannis},
  \bibinfo{person}{Konstantina Papagiannaki}, {and} \bibinfo{person}{Michalis
  Faloutsos}.} \bibinfo{year}{2005}\natexlab{}.
\newblock \showarticletitle{{BLINC: Multilevel Traffic Classification in the
  Dark}}. In \bibinfo{booktitle}{\emph{ACM SIGCOMM}}.
  \bibinfo{pages}{229--240}.
\newblock


\bibitem[\protect\citeauthoryear{Kohno, Broido, and Claffy}{Kohno
  et~al\mbox{.}}{2005}]%
        {kohno2005remote}
\bibfield{author}{\bibinfo{person}{Tadayoshi Kohno}, \bibinfo{person}{Andre
  Broido}, {and} \bibinfo{person}{Kimberly~C Claffy}.}
  \bibinfo{year}{2005}\natexlab{}.
\newblock \showarticletitle{Remote physical device fingerprinting}.
\newblock \bibinfo{journal}{\emph{IEEE Transactions on Dependable and Secure
  Computing}} \bibinfo{volume}{2}, \bibinfo{number}{2} (\bibinfo{year}{2005}),
  \bibinfo{pages}{93--108}.
\newblock


\bibitem[\protect\citeauthoryear{Kotthoff, Thornton, Hoos, Hutter, and
  Leyton-Brown}{Kotthoff et~al\mbox{.}}{2017}]%
        {kotthoff2017auto}
\bibfield{author}{\bibinfo{person}{Lars Kotthoff}, \bibinfo{person}{Chris
  Thornton}, \bibinfo{person}{Holger~H Hoos}, \bibinfo{person}{Frank Hutter},
  {and} \bibinfo{person}{Kevin Leyton-Brown}.} \bibinfo{year}{2017}\natexlab{}.
\newblock \showarticletitle{Auto-WEKA 2.0: Automatic model selection and
  hyperparameter optimization in WEKA}.
\newblock \bibinfo{journal}{\emph{The Journal of Machine Learning Research}}
  \bibinfo{volume}{18}, \bibinfo{number}{1} (\bibinfo{year}{2017}),
  \bibinfo{pages}{826--830}.
\newblock


\bibitem[\protect\citeauthoryear{Lab}{Lab}{[n.\,d.]}]%
        {stratosphere}
\bibfield{author}{\bibinfo{person}{Stratosphere Lab}.}
  \bibinfo{year}{[n.\,d.]}\natexlab{}.
\newblock \bibinfo{booktitle}{\emph{Stratosphere Lab Malware Capture Facility
  Project}}.
\newblock
\urldef\tempurl%
\url{https://www.stratosphereips.org/datasets-malware}
\showURL{%
\tempurl}
\newblock
\shownote{iot malware}.


\bibitem[\protect\citeauthoryear{LeDell and Poirier}{LeDell and
  Poirier}{2020}]%
        {ledell2020h2o}
\bibfield{author}{\bibinfo{person}{Erin LeDell} {and} \bibinfo{person}{S
  Poirier}.} \bibinfo{year}{2020}\natexlab{}.
\newblock \showarticletitle{H2o automl: Scalable automatic machine learning}.
  In \bibinfo{booktitle}{\emph{Proceedings of the AutoML Workshop at ICML}},
  Vol.~\bibinfo{volume}{2020}.
\newblock


\bibitem[\protect\citeauthoryear{Lippmann, Fried, Piwowarski, and
  Streilein}{Lippmann et~al\mbox{.}}{2003}]%
        {lippmann2003passive}
\bibfield{author}{\bibinfo{person}{Richard Lippmann}, \bibinfo{person}{David
  Fried}, \bibinfo{person}{Keith Piwowarski}, {and} \bibinfo{person}{William
  Streilein}.} \bibinfo{year}{2003}\natexlab{}.
\newblock \showarticletitle{Passive operating system identification from
  {TCP/IP} packet headers}. In \bibinfo{booktitle}{\emph{Workshop on Data
  Mining for Computer Security}}, Vol.~\bibinfo{volume}{40}.
\newblock


\bibitem[\protect\citeauthoryear{Lyon}{Lyon}{2009}]%
        {nmap}
\bibfield{author}{\bibinfo{person}{Gordon~Fyodor Lyon}.}
  \bibinfo{year}{2009}\natexlab{}.
\newblock \bibinfo{booktitle}{\emph{Nmap Network Scanning: The Official Nmap
  Project Guide to Network Discovery and Security Scanning}}.
\newblock \bibinfo{publisher}{Insecure}, \bibinfo{address}{USA}.
\newblock
\showISBNx{0979958717}


\bibitem[\protect\citeauthoryear{MacMillan, Holland, and Mittal}{MacMillan
  et~al\mbox{.}}{2020}]%
        {macmillan2020evaluating}
\bibfield{author}{\bibinfo{person}{Kyle MacMillan}, \bibinfo{person}{Jordan
  Holland}, {and} \bibinfo{person}{Prateek Mittal}.}
  \bibinfo{year}{2020}\natexlab{}.
\newblock \showarticletitle{Evaluating Snowflake as an Indistinguishable
  Censorship Circumvention Tool}.
\newblock \bibinfo{journal}{\emph{arXiv preprint arXiv:2008.03254}}
  (\bibinfo{year}{2020}).
\newblock


\bibitem[\protect\citeauthoryear{Miller}{Miller}{2020}]%
        {ttl}
\bibfield{author}{\bibinfo{person}{Tony Miller}.}
  \bibinfo{year}{2020}\natexlab{}.
\newblock \bibinfo{title}{{Passive OS Fingerprinting: Details and Techniques}}.
\newblock \bibinfo{howpublished}{\url{http://www.ouah.org/incosfingerp.htm}}.
\newblock


\bibitem[\protect\citeauthoryear{Mirsky, Doitshman, Elovici, and
  Shabtai}{Mirsky et~al\mbox{.}}{2018}]%
        {kitsune}
\bibfield{author}{\bibinfo{person}{Yisroel Mirsky}, \bibinfo{person}{Tomer
  Doitshman}, \bibinfo{person}{Yuval Elovici}, {and} \bibinfo{person}{Asaf
  Shabtai}.} \bibinfo{year}{2018}\natexlab{}.
\newblock \showarticletitle{Kitsune: an ensemble of autoencoders for online
  network intrusion detection}. In \bibinfo{booktitle}{\emph{Network and
  Distributed System Security Symposium, {NDSS}}}. \bibinfo{address}{San Diego,
  CA, USA}.
\newblock


\bibitem[\protect\citeauthoryear{Nasr, Bahramali, and Houmansadr}{Nasr
  et~al\mbox{.}}{2018}]%
        {10.1145/3243734.3243824}
\bibfield{author}{\bibinfo{person}{Milad Nasr}, \bibinfo{person}{Alireza
  Bahramali}, {and} \bibinfo{person}{Amir Houmansadr}.}
  \bibinfo{year}{2018}\natexlab{}.
\newblock \showarticletitle{DeepCorr: Strong Flow Correlation Attacks on Tor
  Using Deep Learning} \emph{(\bibinfo{series}{CCS '18})}.
  \bibinfo{address}{Toronto, Canada}.
\newblock


\bibitem[\protect\citeauthoryear{Nayak, Misra, and Behera}{Nayak
  et~al\mbox{.}}{2014}]%
        {nayak2014impact}
\bibfield{author}{\bibinfo{person}{SC Nayak}, \bibinfo{person}{Bijan~B Misra},
  {and} \bibinfo{person}{Himansu~Sekhar Behera}.}
  \bibinfo{year}{2014}\natexlab{}.
\newblock \showarticletitle{Impact of data normalization on stock index
  forecasting}.
\newblock \bibinfo{journal}{\emph{International Journal of Computer Information
  Systems and Industrial Management Applications}} \bibinfo{volume}{6},
  \bibinfo{number}{2014} (\bibinfo{year}{2014}), \bibinfo{pages}{257--269}.
\newblock


\bibitem[\protect\citeauthoryear{NetML}{NetML}{[n.\,d.]}]%
        {netmlleaderboard}
\bibfield{author}{\bibinfo{person}{NetML}.}
  \bibinfo{year}{[n.\,d.]}\natexlab{}.
\newblock \bibinfo{booktitle}{\emph{NetML Network Traffic Analytics Challenge
  2020}}.
\newblock
\urldef\tempurl%
\url{https://eval.ai/web/challenges/challenge-page/526/leaderboard/1473}
\showURL{%
\tempurl}
\newblock
\shownote{netML}.


\bibitem[\protect\citeauthoryear{ntop}{ntop}{[n.\,d.]}]%
        {pfring}
\bibfield{author}{\bibinfo{person}{ntop}.} \bibinfo{year}{[n.\,d.]}\natexlab{}.
\newblock \bibinfo{booktitle}{\emph{PF\_RING, High-speed packet capture,
  filtering and analysis}}.
\newblock
\urldef\tempurl%
\url{https://www.ntop.org/products/packet-capture/pf_ring/}
\showURL{%
\tempurl}
\newblock
\shownote{pfring}.


\bibitem[\protect\citeauthoryear{Oh, Sunkam, and Hopper}{Oh
  et~al\mbox{.}}{2019}]%
        {oh2019p1}
\bibfield{author}{\bibinfo{person}{Se~Eun Oh}, \bibinfo{person}{Saikrishna
  Sunkam}, {and} \bibinfo{person}{Nicholas Hopper}.}
  \bibinfo{year}{2019}\natexlab{}.
\newblock \showarticletitle{p1-FP: Extraction, Classification, and Prediction
  of Website Fingerprints with Deep Learning}.
\newblock \bibinfo{journal}{\emph{Proceedings on Privacy Enhancing
  Technologies}} \bibinfo{volume}{2019}, \bibinfo{number}{3}
  (\bibinfo{year}{2019}), \bibinfo{pages}{191--209}.
\newblock


\bibitem[\protect\citeauthoryear{p0f}{p0f}{2016}]%
        {p0f}
p0f \bibinfo{year}{2016}\natexlab{}.
\newblock \bibinfo{title}{{p0f v3 (version 3.09b)}}.
\newblock \bibinfo{howpublished}{\url{http://lcamtuf.coredump.cx/p0f3}}.
\newblock


\bibitem[\protect\citeauthoryear{Padhye and Floyd}{Padhye and Floyd}{2001}]%
        {10.1145/964723.383083}
\bibfield{author}{\bibinfo{person}{Jitendra Padhye} {and}
  \bibinfo{person}{Sally Floyd}.} \bibinfo{year}{2001}\natexlab{}.
\newblock \showarticletitle{On Inferring {TCP} Behavior}.
\newblock \bibinfo{journal}{\emph{SIGCOMM Comput. Commun. Rev.}}
  \bibinfo{volume}{31}, \bibinfo{number}{4} (\bibinfo{date}{Aug.}
  \bibinfo{year}{2001}), \bibinfo{pages}{287–298}.
\newblock
\showISSN{0146-4833}
\urldef\tempurl%
\url{https://doi.org/10.1145/964723.383083}
\showDOI{\tempurl}


\bibitem[\protect\citeauthoryear{Panchenko, Lanze, Pennekamp, Engel, Zinnen,
  Henze, and Wehrle}{Panchenko et~al\mbox{.}}{2016}]%
        {panchenko2016website}
\bibfield{author}{\bibinfo{person}{Andriy Panchenko}, \bibinfo{person}{Fabian
  Lanze}, \bibinfo{person}{Jan Pennekamp}, \bibinfo{person}{Thomas Engel},
  \bibinfo{person}{Andreas Zinnen}, \bibinfo{person}{Martin Henze}, {and}
  \bibinfo{person}{Klaus Wehrle}.} \bibinfo{year}{2016}\natexlab{}.
\newblock \showarticletitle{Website Fingerprinting at Internet Scale.}. In
  \bibinfo{booktitle}{\emph{NDSS}}.
\newblock


\bibitem[\protect\citeauthoryear{Paxson}{Paxson}{1997}]%
        {10.1145/263109.263160}
\bibfield{author}{\bibinfo{person}{Vern Paxson}.}
  \bibinfo{year}{1997}\natexlab{}.
\newblock \showarticletitle{Automated Packet Trace Analysis of {TCP}
  Implementations}.
\newblock \bibinfo{journal}{\emph{SIGCOMM Comput. Commun. Rev.}}
  \bibinfo{volume}{27}, \bibinfo{number}{4} (\bibinfo{date}{Oct.}
  \bibinfo{year}{1997}), \bibinfo{pages}{167–179}.
\newblock
\showISSN{0146-4833}
\urldef\tempurl%
\url{https://doi.org/10.1145/263109.263160}
\showDOI{\tempurl}


\bibitem[\protect\citeauthoryear{Ren, Dubois, and Choffnes}{Ren
  et~al\mbox{.}}{2019}]%
        {crossmarketdataset}
\bibfield{author}{\bibinfo{person}{Jingjing Ren}, \bibinfo{person}{Daniel~J.
  Dubois}, {and} \bibinfo{person}{David Choffnes}.}
  \bibinfo{year}{2019}\natexlab{}.
\newblock \showarticletitle{An International View of Privacy Risks for Mobile
  Apps}.
\newblock  (\bibinfo{year}{2019}).
\newblock
\urldef\tempurl%
\url{https://recon.meddle.mobi/papers/cross-market.pdf}
\showURL{%
\tempurl}


\bibitem[\protect\citeauthoryear{Rimmer, Preuveneers, Ju{\'{a}}rez, van
  Goethem, and Joosen}{Rimmer et~al\mbox{.}}{2018}]%
        {DBLP:conf/ndss/RimmerPJGJ18}
\bibfield{author}{\bibinfo{person}{Vera Rimmer}, \bibinfo{person}{Davy
  Preuveneers}, \bibinfo{person}{Marc Ju{\'{a}}rez}, \bibinfo{person}{Tom van
  Goethem}, {and} \bibinfo{person}{Wouter Joosen}.}
  \bibinfo{year}{2018}\natexlab{}.
\newblock \showarticletitle{Automated Website Fingerprinting through Deep
  Learning}. In \bibinfo{booktitle}{\emph{Network and Distributed System
  Security Symposium, {NDSS}}}. \bibinfo{address}{San Diego, CA, USA}.
\newblock


\bibitem[\protect\citeauthoryear{scikit learn}{scikit learn}{[n.\,d.]a}]%
        {sklearn:aps}
\bibfield{author}{\bibinfo{person}{scikit learn}.}
  \bibinfo{year}{[n.\,d.]}\natexlab{a}.
\newblock \bibinfo{booktitle}{\emph{Computing the Average Precision Score}}.
\newblock
\urldef\tempurl%
\url{https://scikit-learn.org/stable/modules/generated/sklearn.metrics.average_precision_score.html}
\showURL{%
\tempurl}
\newblock
\shownote{average precision score}.


\bibitem[\protect\citeauthoryear{scikit learn}{scikit learn}{[n.\,d.]b}]%
        {sklearn:gs}
\bibfield{author}{\bibinfo{person}{scikit learn}.}
  \bibinfo{year}{[n.\,d.]}\natexlab{b}.
\newblock \bibinfo{booktitle}{\emph{Tuning the hyper-parameters of an
  estimator}}.
\newblock
\urldef\tempurl%
\url{https://scikit-learn.org/stable/modules/grid_search.html}
\showURL{%
\tempurl}
\newblock
\shownote{grid search}.


\bibitem[\protect\citeauthoryear{Sharafaldin, Lashkari, and
  Ghorbani}{Sharafaldin et~al\mbox{.}}{2018}]%
        {sharafaldin2018toward}
\bibfield{author}{\bibinfo{person}{Iman Sharafaldin},
  \bibinfo{person}{Arash~Habibi Lashkari}, {and} \bibinfo{person}{Ali~A
  Ghorbani}.} \bibinfo{year}{2018}\natexlab{}.
\newblock \showarticletitle{Toward generating a new intrusion detection dataset
  and intrusion traffic characterization.}. In
  \bibinfo{booktitle}{\emph{ICISSP}}.
\newblock


\bibitem[\protect\citeauthoryear{Shodan}{Shodan}{2020}]%
        {shodan}
\bibfield{author}{\bibinfo{person}{Shodan}.} \bibinfo{year}{2020}\natexlab{}.
\newblock \bibinfo{title}{Shodan}.
\newblock \bibinfo{howpublished}{\url{https://www.shodan.io/}}.
\newblock


\bibitem[\protect\citeauthoryear{Singh and Singh}{Singh and Singh}{2019}]%
        {singh2019investigating}
\bibfield{author}{\bibinfo{person}{Dalwinder Singh} {and}
  \bibinfo{person}{Birmohan Singh}.} \bibinfo{year}{2019}\natexlab{}.
\newblock \showarticletitle{Investigating the impact of data normalization on
  classification performance}.
\newblock \bibinfo{journal}{\emph{Applied Soft Computing}}
  (\bibinfo{year}{2019}), \bibinfo{pages}{105524}.
\newblock


\bibitem[\protect\citeauthoryear{Sirinam, Imani, Juarez, and Wright}{Sirinam
  et~al\mbox{.}}{2018}]%
        {sirinam2018deep}
\bibfield{author}{\bibinfo{person}{Payap Sirinam}, \bibinfo{person}{Mohsen
  Imani}, \bibinfo{person}{Marc Juarez}, {and} \bibinfo{person}{Matthew
  Wright}.} \bibinfo{year}{2018}\natexlab{}.
\newblock \showarticletitle{Deep Fingerprinting: Undermining Website
  Fingerprinting Defenses with Deep Learning}.
\newblock \bibinfo{journal}{\emph{arXiv preprint arXiv:1801.02265}}
  (\bibinfo{year}{2018}).
\newblock
\showeprint[arxiv]{1801.02265}~[cs.CR]


\bibitem[\protect\citeauthoryear{Smart, Malan, and Jahanian}{Smart
  et~al\mbox{.}}{2000}]%
        {smart2000defeating}
\bibfield{author}{\bibinfo{person}{Matthew Smart}, \bibinfo{person}{G~Robert
  Malan}, {and} \bibinfo{person}{Farnam Jahanian}.}
  \bibinfo{year}{2000}\natexlab{}.
\newblock \showarticletitle{Defeating TCP/IP Stack Fingerprinting.}. In
  \bibinfo{booktitle}{\emph{Usenix Security Symposium}}.
\newblock


\bibitem[\protect\citeauthoryear{Sommer and Paxson}{Sommer and Paxson}{2010}]%
        {sommer2010outside}
\bibfield{author}{\bibinfo{person}{Robin Sommer} {and} \bibinfo{person}{Vern
  Paxson}.} \bibinfo{year}{2010}\natexlab{}.
\newblock \showarticletitle{Outside the closed world: On using machine learning
  for network intrusion detection}. In \bibinfo{booktitle}{\emph{2010 IEEE
  symposium on security and privacy}}. IEEE, \bibinfo{pages}{305--316}.
\newblock


\bibitem[\protect\citeauthoryear{tcpdump}{tcpdump}{[n.\,d.]}]%
        {libpcap}
\bibfield{author}{\bibinfo{person}{tcpdump}.}
  \bibinfo{year}{[n.\,d.]}\natexlab{}.
\newblock \bibinfo{booktitle}{\emph{Man page of PCAP}}.
\newblock
\urldef\tempurl%
\url{https://www.tcpdump.org/manpages/pcap.3pcap.html}
\showURL{%
\tempurl}
\newblock
\shownote{libpcap}.


\bibitem[\protect\citeauthoryear{Trimananda, Varmarken, Markopoulou, and
  Demsky}{Trimananda et~al\mbox{.}}{2020}]%
        {trimananda2020packet}
\bibfield{author}{\bibinfo{person}{Rahmadi Trimananda}, \bibinfo{person}{Janus
  Varmarken}, \bibinfo{person}{Athina Markopoulou}, {and}
  \bibinfo{person}{Brian Demsky}.} \bibinfo{year}{2020}\natexlab{}.
\newblock \showarticletitle{Packet-Level Signatures for Smart Home Devices}. In
  \bibinfo{booktitle}{\emph{Network and Distributed System Security Symposium,
  {NDSS}}}. \bibinfo{address}{San Diego, CA, USA}.
\newblock


\bibitem[\protect\citeauthoryear{van Ede, Bortolameotti, Continella, Ren,
  Dubois, Lindorfer, Choffnes, van Steen, and Peter}{van Ede
  et~al\mbox{.}}{2020}]%
        {van2020flowprint}
\bibfield{author}{\bibinfo{person}{Thijs van Ede}, \bibinfo{person}{Riccardo
  Bortolameotti}, \bibinfo{person}{Andrea Continella},
  \bibinfo{person}{Jingjing Ren}, \bibinfo{person}{Daniel~J Dubois},
  \bibinfo{person}{Martina Lindorfer}, \bibinfo{person}{David Choffnes},
  \bibinfo{person}{Maarten van Steen}, {and} \bibinfo{person}{Andreas Peter}.}
  \bibinfo{year}{2020}\natexlab{}.
\newblock \showarticletitle{{Flowprint: Semi-supervised Mobile-app
  Fingerprinting on Encrypted Network Traffic}}. In
  \bibinfo{booktitle}{\emph{Network and Distributed System Security
  Symposium}}. Internet Society.
\newblock


\bibitem[\protect\citeauthoryear{Venkataraman, Caballero, Poosankam, Kang, and
  Song}{Venkataraman et~al\mbox{.}}{2007}]%
        {venkataraman}
\bibfield{author}{\bibinfo{person}{Shobha Venkataraman}, \bibinfo{person}{Juan
  Caballero}, \bibinfo{person}{Pongsin Poosankam}, \bibinfo{person}{Min Kang},
  {and} \bibinfo{person}{Dawn Song}.} \bibinfo{year}{2007}\natexlab{}.
\newblock \showarticletitle{Fig: Automatic Fingerprint Generation.}. In
  \bibinfo{booktitle}{\emph{Network and Distributed System Security Symposium,
  {NDSS}}}.
\newblock


\bibitem[\protect\citeauthoryear{Wang, Dyer, Akella, Ristenpart, and
  Shrimpton}{Wang et~al\mbox{.}}{2015}]%
        {10.1145/2810103.2813715}
\bibfield{author}{\bibinfo{person}{Liang Wang}, \bibinfo{person}{Kevin~P.
  Dyer}, \bibinfo{person}{Aditya Akella}, \bibinfo{person}{Thomas Ristenpart},
  {and} \bibinfo{person}{Thomas Shrimpton}.} \bibinfo{year}{2015}\natexlab{}.
\newblock \showarticletitle{Seeing through Network-Protocol Obfuscation}. In
  \bibinfo{booktitle}{\emph{Proceedings of the 22nd ACM SIGSAC Conference on
  Computer and Communications Security}} (Denver, Colorado, USA)
  \emph{(\bibinfo{series}{CCS ’15})}. \bibinfo{publisher}{Association for
  Computing Machinery}, \bibinfo{address}{New York, NY, USA},
  \bibinfo{pages}{57–69}.
\newblock
\showISBNx{9781450338325}
\urldef\tempurl%
\url{https://doi.org/10.1145/2810103.2813715}
\showDOI{\tempurl}


\bibitem[\protect\citeauthoryear{Wang, Cai, Nithyanand, Johnson, and
  Goldberg}{Wang et~al\mbox{.}}{2014}]%
        {184463}
\bibfield{author}{\bibinfo{person}{Tao Wang}, \bibinfo{person}{Xiang Cai},
  \bibinfo{person}{Rishab Nithyanand}, \bibinfo{person}{Rob Johnson}, {and}
  \bibinfo{person}{Ian Goldberg}.} \bibinfo{year}{2014}\natexlab{}.
\newblock \showarticletitle{Effective Attacks and Provable Defenses for Website
  Fingerprinting}. In \bibinfo{booktitle}{\emph{23rd {USENIX} Security
  Symposium ({USENIX} Security 14)}}. \bibinfo{publisher}{{USENIX}
  Association}, \bibinfo{address}{San Diego, CA}, \bibinfo{pages}{143--157}.
\newblock
\showISBNx{978-1-931971-15-7}
\urldef\tempurl%
\url{https://www.usenix.org/conference/usenixsecurity14/technical-sessions/presentation/wang_tao}
\showURL{%
\tempurl}


\bibitem[\protect\citeauthoryear{Wang and Goldberg}{Wang and Goldberg}{2017}]%
        {10.5555/3241189.3241296}
\bibfield{author}{\bibinfo{person}{Tao Wang} {and} \bibinfo{person}{Ian
  Goldberg}.} \bibinfo{year}{2017}\natexlab{}.
\newblock \showarticletitle{Walkie-Talkie: An Efficient Defense against Passive
  Website Fingerprinting Attacks}. In \bibinfo{booktitle}{\emph{Proceedings of
  the 26th USENIX Conference on Security Symposium}} (Vancouver, BC, Canada)
  \emph{(\bibinfo{series}{SEC’17})}. \bibinfo{publisher}{USENIX Association},
  \bibinfo{address}{USA}, \bibinfo{pages}{1375–1390}.
\newblock
\showISBNx{9781931971409}


\bibitem[\protect\citeauthoryear{Wang, Sheng, Wang, Zeng, Ye, Huang, and
  Zhu}{Wang et~al\mbox{.}}{2017}]%
        {wang2017hast}
\bibfield{author}{\bibinfo{person}{Wei Wang}, \bibinfo{person}{Yiqiang Sheng},
  \bibinfo{person}{Jinlin Wang}, \bibinfo{person}{Xuewen Zeng},
  \bibinfo{person}{Xiaozhou Ye}, \bibinfo{person}{Yongzhong Huang}, {and}
  \bibinfo{person}{Ming Zhu}.} \bibinfo{year}{2017}\natexlab{}.
\newblock \showarticletitle{HAST-IDS: Learning hierarchical spatial-temporal
  features using deep neural networks to improve intrusion detection}.
\newblock \bibinfo{journal}{\emph{IEEE Access}}  \bibinfo{volume}{6}
  (\bibinfo{year}{2017}), \bibinfo{pages}{1792--1806}.
\newblock


\bibitem[\protect\citeauthoryear{{Wang}, {Zhu}, {Wang}, {Zeng}, and
  {Yang}}{{Wang} et~al\mbox{.}}{2017}]%
        {8004872}
\bibfield{author}{\bibinfo{person}{W. {Wang}}, \bibinfo{person}{M. {Zhu}},
  \bibinfo{person}{J. {Wang}}, \bibinfo{person}{X. {Zeng}}, {and}
  \bibinfo{person}{Z. {Yang}}.} \bibinfo{year}{2017}\natexlab{}.
\newblock \showarticletitle{End-to-end encrypted traffic classification with
  one-dimensional convolution neural networks}. In
  \bibinfo{booktitle}{\emph{2017 IEEE International Conference on Intelligence
  and Security Informatics (ISI)}}. \bibinfo{address}{Beijing, China}.
\newblock


\bibitem[\protect\citeauthoryear{Williams, Zander, and Armitage}{Williams
  et~al\mbox{.}}{2006}]%
        {10.1145/1163593.1163596}
\bibfield{author}{\bibinfo{person}{Nigel Williams}, \bibinfo{person}{Sebastian
  Zander}, {and} \bibinfo{person}{Grenville Armitage}.}
  \bibinfo{year}{2006}\natexlab{}.
\newblock \showarticletitle{A Preliminary Performance Comparison of Five
  Machine Learning Algorithms for Practical {IP} Traffic Flow Classification}.
\newblock \bibinfo{journal}{\emph{SIGCOMM Comput. Commun. Rev.}}
  \bibinfo{volume}{36}, \bibinfo{number}{5} (\bibinfo{date}{Oct.}
  \bibinfo{year}{2006}), \bibinfo{pages}{5–16}.
\newblock
\showISSN{0146-4833}
\urldef\tempurl%
\url{https://doi.org/10.1145/1163593.1163596}
\showDOI{\tempurl}


\bibitem[\protect\citeauthoryear{Yu, Long, and Cai}{Yu et~al\mbox{.}}{2017}]%
        {yu2017network}
\bibfield{author}{\bibinfo{person}{Yang Yu}, \bibinfo{person}{Jun Long}, {and}
  \bibinfo{person}{Zhiping Cai}.} \bibinfo{year}{2017}\natexlab{}.
\newblock \showarticletitle{Network intrusion detection through stacking
  dilated convolutional autoencoders}.
\newblock \bibinfo{journal}{\emph{Security and Communication Networks}}
  \bibinfo{volume}{2017} (\bibinfo{year}{2017}).
\newblock


\end{thebibliography}
\pagebreak

\clearpage
\appendix
\section{Appendix}
\label{sec:appendix}

\subsection{Option Representation Evaluation}
\label{appendix-ordering}

Section~\ref{sec:design} outlines a set of representation requirements and pitfalls of
strawman representations. One issue with a semantic representation that renders it
unusable as a \textit{standard} representation is that option ordering is not
preserved. We now aim to exhibit the performance degradation that can occur
when option ordering is not preserved. 

We set up a classification problem using the dataset fully explained in
Section~\ref{sec:active}. At a high level, the dataset consists of fingerprints
for 15 classes of devices probed with Nmap. We take the TCP responses from each
fingerprint and generate two representations of the TCP options in each packet,
one using \name{}, and one using a semantic representation. For the semantic
representation we
parse all of the options and consider each TCP option as a continuous valued feature, with the
name of the feature being the TCP option and the value being its corresponding
value in the packet. \name{} fingerprints are the bitmap representation of the
options, which preserves ordering.

\begin{table}[h]
    \centering
    \small
\begin{tabular}{@{}lll@{}}
\toprule
Model             & Representation            & F1   \\ \midrule
Catboost          & \multirow{6}{*}{Semantic} & 75.1 \\
ExtraTrees        &                           & 72.9 \\
LightGBM          &                           & 74.4 \\
Neural Network    &                           & 68.2 \\
Random Forest     &                           & 73.6 \\
Weighted Ensemble &                           & 75.6 \\ \midrule
Catboost          & \multirow{6}{*}{nPrint}   & 85.2 \\
Extra Trees       &                           & 83.3 \\
LightGBM          &                           & 83.1 \\
Neural Network    &                           & 82.3 \\
Random Forest     &                           & 83.3 \\
Weighted Ensemble &                           & 85.9 \\ \bottomrule
\end{tabular}
    \caption{Option ordering increases performance across all models.}
    \label{tab:order}
\end{table}

Table~\ref{tab:order} shows the performance degradation that occurs when
ordering is lost across a wide array of models. In general, we see over 10\% increase in F1
scores for \name{} over the semantic representation.


\subsection{nPrintML Example}
\label{appendix-tutorial}

Section~\ref{sec:implement} introduces nPrintML, our pipeline to connect
nPrint and autoML. nPrintML enables researchers with network traffic and code to
perform state-of-the-art traffic analysis without writing code in minutes. Here,
we present an example of reproducing our results in Section~\ref{sec:app}
\textit{from scratch}. 

\begin{enumerate}
    \item Clone the repository to get the data.
    \item Uncompress the traffic traces.
    \item Write a small script to generate labels. In cases where we have labels
        apirori, \textbf{no code} needs to be written.
    \item Run the label generation script
    \item run nPrintML to perform traffic analysis.
\end{enumerate}

These steps, instantiated.

\begin{lstlisting}
# 1. clone
$ git clone repo_name
# 2. uncompress
$ tar -xvf dataset.tar.gz
\end{lstlisting}
\begin{lstlisting}[language=Python]
# 3. Generate Labels, not necessary if labels exist
import sys
import pathlib

# Example file name: dataset/facebook/windows_chrome_facebook_1383.pcap 
# Get file paths
paths = list(pathlib.Path(sys.argv[1]).rglob('*.pcap'))
for path in paths:
    # Build label
    tokens = str(path.stem).split('_')
    label = '{0}_{1}'.format(tokens[0], tokens[1])
    print('{0},{1}'.format(path, label)) 
\end{lstlisting}

\begin{lstlisting}    
# 4. Generate Labels
$ python gen_labels.py dataset/ > labels.txt
\end{lstlisting}

\begin{lstlisting}
# 5. run nPrintML with different configurations

# IPv4, UDP, first 10 payload bytes of each packet:
$ nprintML -L labels.txt -a pcap \
--pcap_dir dataset/ -4 -u -p 10 

# First 100 payload bytes of each packet:
$ nprintML -L labels.txt -a pcap \
--pcap_dir dataset/ -p 100 

# UDP headers only:
$ nprintML -L labels.txt -a pcap  \
--pcap_dir dataset/ -u
\end{lstlisting}

\begin{table*}[t]
\centering
\begin{tabular}{@{}llr@{}}
    \toprule
    Test Name & Summary & Nmap Weight \\ \midrule
Explicit Congestion Notification & TCP Explicit Congestion control flag. & 100 \\
\rowcolor{Gray}
ICMP Response Code & ICMP Response Code. & 100 \\
Integrity of returned probe IP Checksum & Valid checksum in an ICMP port unreachable. & 100 \\
\rowcolor{Gray}
Integrity of returned probe UDP Checksum & UDP header checksum received match. & 100 \\
IP ID Sequence Generation Algorithm & Algorithm for IP ID. & 100 \\
\rowcolor{Gray}
IP Total Length & Total length of packet. & 100 \\
Responsiveness & Target responded to a given probe. & 100 \\
\rowcolor{Gray}
Returned probe IP ID value & IP ID value. & 100 \\
Returned Probe IP Total Length & IP Length of an ICMP port unreachable. & 100 \\
\rowcolor{Gray}
TCP Timestamp Option Algorithm & TCP timestamp option algorithm. & 100 \\
Unused Port unreachable Field Nonzero & Last 4 bytes of ICMP port unreachable message not zero. & 100 \\
\rowcolor{Gray}
Shared IP ID Sequence Boolean & Shared IP ID Sequence between TCP and ICMP. & 80 \\
TCP ISN Greatest Common Divisor & Smallest TCP ISN increment. & 75 \\
\rowcolor{Gray}
Don’t Fragment ICMP & IP Don’t Fragment bit for ICMP probes. & 40 \\
TCP Flags & TCP flags. & 30 \\
\rowcolor{Gray}
TCP ISN Counter Rate & Average rate of increase for the TCP ISN. & 25 \\
TCP ISN Sequence Predictability Index & Variability in the TCP ISN. & 25 \\
\rowcolor{Gray}
IP Don’t Fragment Bit & IP Don’t Fragment bit. & 20 \\
TCP Acknowledgment Number & TCP acknowledgment number. & 20 \\
\rowcolor{Gray}
TCP Miscellaneous Quirks & TCP implementations, e.g, reserved field in TCP header. & 20 \\
TCP Options Test & TCP header options, preserving order. & 20  \\
\rowcolor{Gray}
TCP Reset Data Checksum & Checksum of data in TCP reset packet. & 20 \\
TCP Sequence Number & TCP sequence number. & 20 \\
\rowcolor{Gray}
IP Initial Time-To-Live & IP initial time-to-live. & 15 \\
TCP Initial Window Size & TCP window size. & 15 \\ \bottomrule
\end{tabular} 

\caption{Nmap's highly complex device detection tests, which are used to
    generate a fingerprint for each device.} 
\label{tab:nmap-tests}
\end{table*}

\newpage

\subsection{Nmap Tests}\label{appendix-nmap}


\onecolumn
\begin{lstlisting}
  -4, --ipv4                 include ipv4 headers
  -6, --ipv6                 include ipv6 headers
  -A, --absolute_timestamps  include absolute timestmap field
  -c, --count=INTEGER        number of packets to parse (if not all)
  -C, --csv_file=FILE        csv (hex packets) infile
  -d, --device=STRING        device to capture from if live capture
  -e, --eth                  include eth headers
  -f, --filter=STRING        filter for libpcap
  -F, --fill_int=INT8_T      integer to fill missing bits with
  -h, --nprint_filter_help   print regex possibilities
  -i, --icmp                 include icmp headers
  -N, --nPrint_file=FILE     nPrint infile
  -O, --write_index=INTEGER  Output file Index (first column) Options:
                             0: source IP (default)
                             1: destination IP
                             2: source port
                             3: destination port
                             4: flow (5-tuple)
  -p, --payload=PAYLOAD_SIZE include n bytes of payload
  -P, --pcap_file=FILE       pcap infile
  -R, --relative_timestamps  include relative timestamp field
  -S, --stats                print stats about packets processed when finished
  -t, --tcp                  include tcp headers
  -u, --udp                  include udp headers
  -V, --verbose              print human readable packets with nPrints
  -W, --write_file=FILE      file for output, else stdout
  -x, --nprint_filter=STRING regex to filter bits out of nPrint. nprint -h for
                             details
  -?, --help                 Give this help list
      --usage                Give a short usage message
      --version              Print program version
\end{lstlisting}

\subsection{nPrint Configuration Options}\label{appendix:config}

\end{document}